\documentclass[epj]{svjour}
 \usepackage{epsfig}
 \usepackage{amsmath}
 \usepackage{amssymb}
 \allowdisplaybreaks[2]
\newcommand{\aop}{\hat{a}^{\mbox{}}}
\newcommand{\aopk}{\hat{a}^{\dagger}}
\newcommand{\alop}{\hat{\alpha}^{\mbox{}}} 
\newcommand{\alopk}{\hat{\alpha}^{\dagger}}

\newcommand{\Nop}{\hat N}

\newcommand{\bra}[1]{ \langle #1 |} 
\newcommand{\ket}[1]{ | #1 \rangle} 
 
\newcommand{\kBCS}{|{\rm BCS} \rangle} 
 
\newcommand{\kvac}{| 0 \rangle}

\newcommand{\rmd}{{\rm d}} 
\newcommand{\iunit}{{\rm i}} 
\newcommand{\vnabla}{{\bf \nabla}} 
 
\newcommand{\gapfive}{\Delta^{(5)}} 
\newcommand{\gapfour}{\Delta^{(4)}} 
\newcommand{\gapthree}{\Delta^{(3)}} 
\newcommand{\gapb}{\Delta^{({\rm b})}} 
\newcommand{\gapuv}{\langle uv  \Delta \rangle} 
\newcommand{\gapvv}{\langle v^2 \Delta \rangle} 
\newcommand{\Equasi}{E_{\rm quasi}} 
 
\newcommand{\rvec}{\vec{r}} 
\newcommand{\rvecp}{\vec{r}'}

\newcommand{\xvec}{\vec{x}} 
\newcommand{\xvecp}{\vec{x}'}

\newcommand{\trace}{{\rm tr}} 
\newcommand{\EF}{{\cal E}}
\begin{document} 
 
\title{Pairing Gaps from Nuclear Mean--Field Models} 
\author{M. Bender\inst{1,2,3} \and 
        K. Rutz\inst{1} \and 
        P.--G. Reinhard\inst{4,5} \and 
        J. A. Maruhn\inst{1,5} 
} 
 
\institute{Institut f\"ur Theoretische Physik, 
           Universit\"at Frankfurt, 
           Robert--Mayer--Str.~10, 
           D--60325 Frankfurt am Main, Germany. 
           \and 
           Department of Physics and Astronomy, 
           The University of North Carolina, 
           Chapel Hill, NC 27516, U.S.A. 
           \and 
           Department of Physics and Astronomy, 
           The University of Tennessee, 
           Knoxville, TN 37996, U.S.A. 
           \and 
           Institut f\"ur Theoretische Physik II, 
           Universit\"at Erlangen--N\"urnberg, 
           Staudtstr.~7, 
           D--91058 Erlangen, Germany. 
           \and 
           Joint Institute for Heavy--Ion Research, 
           Oak Ridge National Laboratory, 
           P.~O.~Box 2008, Oak Ridge, TN 37831, U.S.A. 
} 

\date{May 10 2000}
%
%======================================================================= 
%
\abstract{
We discuss the pairing gap, a measure for nuclear pairing correlations, 
in chains of spherical, semi--magic nuclei
in the framework of self--consistent nuclear mean--field 
models. The equations for the conventional BCS model and the approximate 
projection--before--variation Lipkin--Nogami method are formulated in 
terms of local density functionals for the effective interaction. 
We calculate the Lipkin--Nogami corrections of both the mean--field 
energy and the pairing energy. Various definitions of the pairing 
gap are discussed as three--point, four--point and five--point 
mass--difference formulae, averaged matrix elements of the pairing potential, 
and single--quasiparticle energies. Experimental values for the pairing 
gap are compared with calculations employing both a delta pairing force 
and a density--dependent delta interaction in the BCS and Lipkin--Nogami 
model. Odd--mass nuclei are calculated in the spherical blocking 
approximation which neglects part of the the core polarization in 
the odd nucleus. We find that the five--point mass difference formula 
gives a very robust description of the odd--even staggering, other 
approximations for the gap may differ from that up to $30\%$ for 
certain nuclei.
}

\maketitle 
% 
%======================================================================= 
% 
\section{Introduction} 
Pair correlations, which play a crucial role in superconducting solids 
\cite{Bar57a}, also constitute an important complement of nuclear shell 
structure \cite{Boh58a,Bel58a}. Most often, pairing is treated in the 
so--called BCS approximation \cite{EGIII,Ringbook,Nilssonbook}, which 
was introduced in the original paper of Bardeen, Cooper and Schrieffer 
\cite{Bar57a}.  The nuclear applications are particular in two 
respects: first, nuclei are finite, in fact rather small, objects, and 
second, we do not yet have a sufficiently reliable microscopic nuclear 
many--body theory from which we could deduce the appropriate pairing 
interaction and its strength. The second problem causes two further 
questions. First, one needs to develop a reliable and manageable form 
for the pairing energy functional, and second, one has to determine an 
appropriate pairing strength for a given functional.

There exist various prescriptions for the pairing 
energy functional. Schematic pairing forces basically  
consist of defining a small band of pairing--active states and 
parameterizing one typical pairing matrix element in dependence on the 
system size, i.e.\ neutron and proton number. They are convenient and 
successful in many respects. But they are plagued by serious 
disadvantages: the coupling to continuum states is much exaggerated 
and the parameterization in terms of system size becomes questionable  
for deformed systems along the fission path. To avoid these  
problems local two--body pairing forces are increasingly used 
which is particularly 
satisfying in connection with self--consistent mean--field calculations  
\cite{Ton79a,Kri90a,Dob95a,Dob96c}. There are even some 
mean--field models like the Gogny forces \cite{Dob96c,Gogrev} or the 
particular Skyrme force SkP \cite{Dob84a} which aim at a simultaneous 
description of the particle--particle and particle--hole channels
of the effective interaction using the same force.  
That is not a necessary condition. The simple local pairing energy  
functionals also provide a very good description of 
pairing properties throughout the whole chart of isotopes using 
only two universal strength parameters, one for protons and one for 
neutrons. This makes these forms for the pairing energy functional 
more reliable for calculations of deformed nuclei and of nuclei far 
off the valley of stability than the widely used schematic pairing 
forces. In the following we will concentrate on the class of local 
pairing interactions.

The small particle number of nuclei often interferes with the fact that  
the BCS ground state mixes particle numbers with a relative spread of 
order $1/\sqrt{N}$. In principle, one has to perform a projection of the 
BCS state before varia\-tion, but this exact particle number 
projection can be very cumbersome, see, e.g., \cite{Mang,MONSTER}.  
The Lipkin--Nogami (LN) scheme offers a reliable approximate projection  
method \cite{Lip60a,Nog64a,Pra73a,Flo}. It 
has the technical advantage that it is formulated completely in terms 
of BCS expectation values which makes its numerical implementation 
very simple, for detailed discussion see \cite{Pra73a,Flo,Zhe92a,Dob92a}.
We will discuss the LN scheme side by side with  the BCS treatment.

There remains as the last and crucial problem the determination  
of an appropriate pairing strength.  Insufficient microscopic information 
requires that one recurs to a phenomenological assessment.  This line 
of development has been followed with increasing accuracy since the 
introduction of pairing in nuclei, for a comprehensive 
compilation of pairing with schematic forces see \cite{Mol92a}.  Also 
the local two--body pairing interactions leave the overall strength as a 
free parameter to be fixed phenomenologically.  This has to be 
done with respect to an observable sensitive to pairing correlations. Such 
an observable would ideally be provided by the pairing gap which, 
however, is not directly accessible in experiments.  The observable 
quantity with probably closest relation to the pairing gap is the 
pronounced odd--even mass staggering of nuclei, which is usually 
used for fitting the pairing strength. Even here, though, there remains 
a choice of several recipes on how to extract the pairing gap or 
odd--even staggering, respectively, from a combination of neighboring 
mass values. From the theoretical side, of course, one has more direct 
access to a gap. But even there ambiguities emerge and one is left  
with several possibilities to define measures for the pairing 
correlations.

It is the aim of this paper to compare and discuss the various 
definitions of a pairing gap in models of local two-body pairing 
interactions in connection with self--consistent mean--field  
approaches, both at the level of the conventional BCS approach as 
well as of the LN method. Furthermore, we will compare two variants of 
the local zero--range pairing forces, a delta force and a 
density--dependent form \cite{Dob95a,Dob96c}. Last but not least, the 
discussion will extend to exotic nuclei where differences between the 
various definitions for the pairing gap and options for the pairing 
method become particularly obvious.

The paper is outlined as follows: In Section~\ref{sect:frame} we  
present the basic equations of the Lipkin--Nogami model employing  
local interactions in the framework of the Skyrme--Hartree--Fock model 
needed for our discussion. 
In Section~\ref{sect:gap} various approximations for the pairing gap  
are discussed, in Section~\ref{sect:results} the results are presented, 
Section~\ref{sect:conclusions} summarizes our findings. 
% 
%======================================================================= 
% 
\section{The Theoretical Framework} 
\label{sect:frame} 
We investigate the pairing gap in the framework of self--consistent 
mean--field models. Pairing correlations are treated on the HF+BCS level 
where the equations of motion are derived by independent variation
with respect to single--particle wave functions and occupation amplitudes.
This is a widely used approximation to the more involved HFB approach 
where wave functions and occupation amplitudes are varied simultaneously 
\cite{Ringbook}. The BCS approximation 
is applicable for all well--bound nuclei, i.e.\ for most of the nuclei 
discussed throughout this paper. It becomes critical only for nuclei 
close to the neutron or proton drip--line, see e.g.\ \cite{Dob84a},
and such nuclei are not considered here.  All conclusions about pairing 
gaps drawn in this paper can thus safely be derived in
the BCS approximation. 
%
%----------------------------------------------------------------------
% 
\subsection{The Mean Field} 
\label{subsect:meanfield} 
Presently the most widely used self--consistent mean--field models 
are the non--relativistic Hartree--Fock approach
with either the Skyrme (SHF) \cite{SHFrev} or the Gogny force 
\cite{Gogrev}, and the relativistic mean--field model \cite{RMFrev}.  
They all provide a well-adjusted effective energy functional  
for nuclear mean--field calculations.  
We choose the SHF model for the present investigation. The description  
starts from an energy functional 
\begin{equation} 
\label{eq:Efu} 
{\cal E} 
= {\cal E}_{\rm mf} 
  + {\cal E}_{\rm pair} 
\quad , 
\end{equation} 
whose mean--field part ${\cal E}_{\rm mf}$ is formulated in terms of the  
local distributions of density $\rho$, kinetic density $\tau$, and  
spin--orbit current ${\bf J}$.  
At this point it is not necessary to unfold all details of this  
rather elaborate functional, we abbreviate the dependence with the most  
general case, the full one--body density matrix 
\begin{equation} 
\label{eq:densdef} 
\hat{\rho} 
\equiv \rho(\xvec, \xvecp) 
=      \langle \hat{\psi}^\dagger (\xvecp) \hat{\psi} (\xvec) \rangle 
\end{equation} 
from which all local densities and currents can be derived, see 
appendix~\ref{Subesct:LRskyrme} for details. $\hat{\psi}^\dagger(\xvec)$  
creates a particle with spin projection $\sigma/2$ at the  
space point $\vec{r}$. 
Throughout this paper \mbox{$\langle \cdots \rangle$} denotes BCS  
expectation values. For the Skyrme energy functional we choose the  
rather recent parameterization SkI4 \cite{Rei95a}. The pairing energy  
functional depends additionally on the local pair density $\chi$
as introduced in Section \ref{subsect:pairfu}.
Variation of the energy functional  ${\cal E}$ with respect to the 
single--particle wave functions $\phi_k$ yields the mean--field equations
\begin{equation} 
\label{eq:mfH} 
\hat{h} \, \phi_k = \varepsilon_k \, \phi_k 
\quad \mbox{with} \quad 
\hat{h} = \frac{\delta {\cal E}}{\delta\hat{\rho}} 
\quad , 
\end{equation} 
where we have neglected the contributions from the variation of the  
pairing density $\chi$ in the energy functional to the 
equations--of--motion of the single--particle states $\phi_k$. This 
constitutes the BCS approximation to pairing (see \cite{Rei97a} 
for the discussion of the full HFB equations in the representation 
in natural orbitals that is used here). The mean--field equations
are solved on a grid in coordinate space with the damped gradient
iteration method and a Fourier representation of the derivatives.
The numerical techniques are summarized in \cite{Blum}. 
%
%-------------------------------------------------------------------------
%
\subsection{The Pairing Energy Functional}
\label{subsect:pairfu}
We parameterize the effective pairing interaction in terms of a local pairing
energy functional of the form
\begin{equation}
\label{eq:PairFunc}
{\cal E}_{\rm pair}
= \frac{1}{4} \sum_{q\in\{p,n\}}
  \int \! {\rm d}^3r \; \chi_q^* (\textbf{r}) \; \chi_q (\textbf{r}) \;
  G_q(\textbf{r})
\quad ,
\end{equation}
which allows for a spatial modulation of the strength $G(\textbf{r})$. 
$\chi (\textbf{r})$ is the local part of the pair density matrix
\begin{eqnarray}
\label{eq:PairDensity}
\chi_q (\vec{r})
& = & \sum_{\sigma = \pm} \chi_q (\vec{r}, \sigma; \vec{r}, \sigma)
  = - \sum_{\sigma = \pm} \sigma \;
      \langle \hat\psi_q (\vec{r}, - \sigma)
              \hat\psi_q (\vec{r}, \sigma) \rangle
      \nonumber \\
& = & - 2 \sum_{k \in \Omega_q \atop k > 0} u_k \, v_k \,
      | \phi_k (\textbf{r}) |^2
\quad ,
\end{eqnarray}
with $q\in\{p,n\}$.
The $\phi_k$ are the single--particle wave functions and $v_k$,
\mbox{$u_k = \sqrt{1-v_k^2}$} the pairing amplitudes.
We restrict ourselves to stationary states
of time--reversal invariant systems and pairing between like particles only.
This sorts the single--particle states into conjugate pairs
\mbox{$k \leftrightarrow \bar{k} \equiv - k$} with the same spatial
properties but opposite projection of the total angular momentum and
allows to restrict the summation in the pair density to \mbox{$k>0$}.
The time--reversal symmetry also renders the pair density real,
i.e.\ \mbox{$\chi^* ({\mbox{\bf r}}) = \chi ({\mbox{\bf r}})$}.

Two models for the spatial modulation of the pairing strength are
considered here
\begin{equation}
\label{eq:surfforc}
G_q ({\textbf{r}})
= \left\{
  \begin{array}{lll}
   V_{0,q} & \qquad & \mbox{DF,} \\
   V_{0,q} \; \bigg[ 1 - \left( {\displaystyle \frac{\rho({\mbox{\bf r}})}
                                                    {\rho_0} }
                          \right)^\gamma \bigg]
           & \qquad & \mbox{DDDI.}
  \end{array}
  \right.
\end{equation} 
The simpler case (DF) can be deduced from a \emph{delta force} 
for the pairing interaction \cite{Ton79a,Kri90a,Dob95a},
\mbox{$V_{\rm pair}(\textbf{r},\textbf{r}') 
= V_{0,q}\;\delta(\textbf{r}-\textbf{r}')$},
while the other (DDDI) corresponds to a \emph{density--dependent delta 
interaction} \cite{Dob95a,Taj93b,Fay94a}. The additional parameters of 
the DDDI force are set here to \mbox{$\gamma = 1$} and \mbox{$\rho_0 = 
0.16 \; {\rm fm}^{-3}$} (i.e.\ the saturation density of symmetric 
nuclear matter).  More general choices are conceivable 
\cite{Dob84a,Fay96a} but very hard to adjust phenomenologically. 
Thus we keep to the simplest choice above. 
Note that a separate pairing strength $V_{0,p}$ or $V_{0,n}$
is associated to each nucleon sort. This explicit breaking of the 
isospin symmetry in the pairing energy functional 
is standard in nearly all pairing forces and schematic 
models, see e.g.\ \cite{Mol92a,Mad88a}.

Although the pairing matrix elements deduced from the pairing 
functional (\ref{eq:PairFunc}) suppress the contribution 
from unbound states located outside the nucleus considerably  
(as compared to the schematic pairing force), the 
implicit zero--range nature of the pairing force still tends to 
overestimate the coupling to continuum states.  This defect can 
be cured to some extent using finite--range forces like the Gogny force 
\cite{Dob96c,Gogrev} which are, however, cumbersome to handle. We 
prefer to simulate the effect of finite range by introducing smooth 
energy--dependent cutoff weights \cite{Kri90a} 
\begin{equation} 
\label{eq:cutoff} 
f_k 
= \frac{1}{1 + \exp [(\epsilon_k - \lambda_q - \Delta E_q) / \mu_q] } 
\end{equation} 
in the evaluation of the local pair density  
\begin{equation} 
\label{eq:PairDensity2} 
\chi_q (\textbf{r}) 
\Rightarrow 
  - 2 \sum_{k \in \Omega_q \atop k > 0} f_k \, u_k \, v_k \, 
  | \phi_k (\textbf{r}) |^2  
\quad . 
\end{equation} 
The cutoff parameters $\Delta E_q$ and $\mu_q = \Delta E_q/10$ are chosen  
self--adjusting to the actual level density in the vicinity of the Fermi  
energy. $\Delta E_q$ is fixed from the condition that the sum of the  
cutoff weights includes approximately one additional shell of  
single-particle states above the Fermi surface 
\begin{equation} 
\sum_{k \in \Omega_q} f_k 
= N_q + 1.65 \, N_q^{2/3} 
\quad . 
\end{equation} 
% 
%------------------------------------------------------------------------- 
% 
\subsection{The Lipkin--Nogami Equations} 
\label{subsect:paireq} 
 
The LN scheme serves as an approximation to particle--number projected 
BCS. It can be derived by a momentum expansion of the projected BCS 
equations \cite{Lip60a,Nog64a,Pra73a,Flo}. At the end this boils down to 
adding overlaps with the variance of the particle number 
$(\Delta \hat{N}_q)^2$ at various places.

In most cases, the LN method is used with a simple schematic pairing 
interaction in the framework of macro\-sco\-pic--micro\-sco\-pic models and 
self--consistent models for ground states and potential energy surfaces 
\cite{Mol92a,Ben89a,Naz90a,Que90a,Taj92a,Hee93a,Ska93,Mag95a,Rei96a} 
as well as high--spin states 
\cite{Mag93a,Sat94a,Sat94b,Gal94a,Wys95a,Hee95a,Ter95}.  Only 
recently, the LN scheme was employed for a local delta pairing force 
\cite{Cwi96a} and the Gogny force \cite{Val96a,Val97a}. Usually 
only the correction of the pairing energy is calculated; but in
self--consistent models the contribution of the mean field to
the total binding energy is calculated from the BCS state as well, 
so that the correction of the pairing energy has to be complemented 
by a correction of the mean--field energy as considered in 
\cite{Rei96a,Val96a,Val97a}. In this paper, we present and employ 
the LN equations in the context of self--consistent mean--field 
models and for the case of local pairing energy functionals. As 
pairing gaps are the theme of this paper, particular emphasis is 
laid on the properties of the LN scheme relevant for the discussion 
of pairing gaps.

The LN equations are derived by variation of 
\begin{equation} 
\label{eq:calK} 
{\cal K} 
= {\cal E}  
  - \sum_{q \in p,n} \Big(   \lambda_{1,q} \langle \hat{N}_q   \rangle 
                           + \lambda_{2,q} \langle \hat{N}_q^2 \rangle 
                     \Big) 
\quad . 
\end{equation} 
Variation of (\ref{eq:calK}) with respect to the occupation 
amplitudes $v_k$ leads to
\begin{equation} 
\label{eq:occupation} 
v_k^2  
= \frac{1}{2} \left[ 1 - \frac{\epsilon_k^\prime - \lambda_q} 
                              {\sqrt{ (\epsilon_k^\prime - \lambda_q)^2  
                                      + f_k^2 \; \Delta_k^2}}  
\right] 
\quad . 
\end{equation} 
This is the standard expression for the occupation number 
$v_k^2$ in the BCS model \cite{EGIII,Ringbook,Nilssonbook}, here  
containing a state--dependent single--particle gap $\Delta_k$  
times the cutoff factor $f_k$ and a generalized Fermi energy 
\begin{equation}
\label{eq:Efermi}
\lambda_q
= \lambda_{1,q} + 4 \lambda_{2,q} (N_q + 1)
\quad ,
\end{equation}
which is determined from a constraint on particle number. The quantity
$\epsilon_k^\prime$ is a renormalized single--particle energy
\begin{equation}
\label{eq:Erenorm}
\epsilon_k^\prime
= \epsilon_k + 4 \, \lambda_{2,q} v_k^2
\quad .
\end{equation}
In case of time-reversal invariance, the
state--dependent single--particle gaps are given by
\begin{equation}
\label{eq:spGap}
\Delta_k
= \int \! {\rm d}^3r \; \phi_k^\dagger (\textbf{r})
                        \Delta_q (\textbf{r})
                        \phi_k (\textbf{r})
\quad ,
\end{equation}
i.e.\ they are matrix elements of the local pair potential
\begin{equation}
\Delta_q (\textbf{r})
= \frac{\delta {\cal E}_{\rm pair}}{\delta \chi_q (\textbf{r})}
= \frac{1}{2} \; \chi_q (\textbf{r}) \; G_q (\textbf{r})
\quad .
\end{equation}
Note that $\lambda_2$ is not a Lagrange parameter \cite{Flo,Que90a}.
It is held fixed during the variation and is determined after
variation from the additional condition \cite{Pra73a}
\begin{equation}
\label{eq:lambda2}
\lambda_{2,q}
= \frac{\langle (\hat{H}_{\rm mf}+\hat{H}_{\rm pair})
                (\Delta \hat{N}_{2,q})^2 \rangle}
       {\langle \hat{N}_q (\Delta \hat{N}_{2,q})^2 \rangle}
\quad .
\end{equation}
For simplicity of the presentation, Eq.~(\ref{eq:lambda2}) is written for
the case of an underlying many--body Hamiltonian $\hat{H}$. The discussion 
of the more general case of an energy functional used in this paper is 
presented in Appendix \ref{Sect:lambda2}. $\hat{N}_{2,q}$ is the part of 
the particle--number operator that projects onto two--quasiparticle states
\begin{equation}
\label{eq:N_2}
\hat{N}_{2,q}
= \sum_{k \in \Omega_q}
  u_k v_k \, (  \hat\alpha^\dagger_k \hat\alpha^\dagger_{\bar{k}}
              + \hat\alpha_{\bar{k}} \hat\alpha_k )
\quad ,
\end{equation}
while \hbox{$\Delta \hat{N}_{2,q}
= \hat{N}_{2,q} - \langle \hat{N}_{2,q} \rangle$}.
The numerator of Eq.~(\ref{eq:lambda2}) contains, besides the familiar 
contribution from the pairing functional, an additional one from the 
linear response of the mean--field to the particle--number projection, 
see \cite{Rei96a}. The total binding energy after approximate 
particle--number projection is given by
\begin{equation} 
E^{\rm LN} 
= {\cal E}  
  - \sum_{q \in {\rm p,n}} \lambda_{2,q} \;
    \langle (\Delta \Nop_{2,q})^2 \rangle 
\quad . 
\end{equation}
For arbitrary one--body operators the LN expectation value can be calculated
introducing effective LN occupation numbers and local densities, see
\cite{Que90a,Rei96a}.

Thus far the presentation applies to the more involved LN method. The 
BCS approximation is recovered simply by setting
\mbox{$\lambda_{2,q}=0$} in the above equations.
%
%-------------------------------------------------------------------------
%
\subsection{Blocking}
\label{subsect:blocking}
The evaluation of the odd--even staggering involves also nuclei with
odd mass number where one pair of conjugate states has to be blocked, 
i.e.\ taken out of the pairing scheme. One of the blocked states has 
the occupation \mbox{$v_k = 1$}, the other \mbox{$v_{\bar{k}} = 0$}.
In the standard textbook approach the blocked many--body state is 
constructed non--self--consistently from the BCS ground state as a 
one--quasiparticle excitation, see e.g.\ \cite{EGIII,Ringbook,Nilssonbook}
and Section~\ref{Subsect:OQPE}. The generalization of this approach 
for energy functionals is outlined in Appendix~\ref{Sect:OQPE}.

In a self--consistent approach to the blocked many--body state the 
single--particle wave functions and occupation amplitudes have to be 
determined from a variational principle. Blocking a pair of states 
changes the density matrix, Eq.~(\ref{eq:densdef}) and with that the 
single--particle Hamiltonian, Eq.~(\ref{eq:mfH}). Besides a rearrangement 
of the single--particle wave functions the unpaired nucleon causes a 
polarization of the core by breaking rotational and time--reversal 
invariance in the intrinsic frame (we assume spherical 
BCS ground states only), see e.g.\ \cite{ugpap}.
This requires a deformed calculation of the odd--mass nucleus
considering also time--odd contributions to the single--particle
Hamiltonian (\ref{eq:mfH}) as discussed in \cite{ugpap,Dob95c}.

The change in binding energy due to the core polarization depends 
on the properties of the time--odd spin and spin--isospin 
channels of the effective interaction which are not yet well adjusted 
for current mean--field models, see e.g.\ \cite{Eng99a} and references
therein. Calculations with the currently available models suggest that 
the polarization is non--negligible for the description of the odd--even 
staggering \cite{ugpap,Sat98a,Sat98b,Xu99a,uggap}, but effective
interactions with properly adjusted spin and spin--isospin channels
are needed before the effect can be treated quantitatively.
In view of these
uncertainties we restrict ourselves here to the simpler spherical
blocking approximation, where one replaces the blocked
single--particle state by an average over the degenerate states in its 
$j$ shell, restoring rotational and time--reversal invariance of the 
many--body system in the intrinsic frame. In practice this means
that the weight of the blocked $j$ shell is given by $(2j-1) u v$
when calculating the pair density and $(2j-1) v^2 + 1$ when
calculating the local densities and currents. All other states
enter with their full degeneracy $2j+1$. This approximation includes 
the large part of the rearrangement effects 
from monopole polarization, but omits the polarization effects 
from multipole deformations and time--odd currents.

Owing to the rearrangement effects blocking of the single--particle 
state with smallest quasiparticle energy (as defined in 
Sect.~\ref{Subsect:OQPE}) does not necessarily lead to the largest 
possible total binding energy. Therefore one has
to perform calculations for a number of blocked single--particle  
states around the Fermi energy and search for the configuration  
giving the largest total binding energy. 
% 
%======================================================================= 
% 
\section{The Pairing Gap} 
\label{sect:gap} 
 
\subsection{Nuclear Masses and Odd--Even Staggering} 
\label{Subsect:gapb}
The key feature of pairing correlations is the occurrence of an energy 
gap in the excitation spectrum. This gap manifests itself 
in two different kinds of energetic observables: First, there is a  
gap in the quasiparticle excitation spectra of even--even 
nuclei, which does not appear in the spectra of odd--mass number  
or odd--odd nuclei, and second, there occurs a shift between  
the interpolating curves of the ground--state binding energies of 
even--even as compared to odd--mass nuclei, which is called the odd--even 
mass staggering. Usually, the second phenomenon is exploited to define 
the experimental pairing gaps assuming \cite{Mad88a} 
\begin{eqnarray} 
\label{eq:GapShift} 
E_{\rm even-even }(Z,N)  
& = & E_0 (Z,N)                       \quad , \nonumber \\ 
E_{{\rm odd} \; Z}(Z,N)  
& = & E_0 (Z,N) + \Delta_{\rm p}(Z,N) \quad , \\ 
E_{{\rm odd} \; N}(Z,N)  
& = & E_0 (Z,N) + \Delta_{\rm n}(Z,N) \quad . \nonumber 
\end{eqnarray} 
In odd--odd nuclei there is additionally the residual 
interaction between the unpaired proton and neutron, but this case 
will not be considered in the present discussion. 
In a self--consistent mean--field approach $E_0$ is the (negative) 
energy  of the fully paired many--particle wave function, i.e.\ the
BCS ground state, while $\Delta_q$ is the energy lost due to the 
blocking of a pair of conjugate states by the odd nucleon. The gap 
introduced with (\ref{eq:GapShift}) has to be interpreted carefully.
The gap as defined through the separation (\ref{eq:GapShift}) 
contains more than pure pairing correlations. It includes unavoidably 
all polarization effects from the mean field as outlined in 
Sect.~\ref{subsect:blocking}. Despite of these uncertainties 
(\ref{eq:GapShift}) provides the starting point for the definition 
of experimentally accessible pairing gaps which will be discussed 
in Sect.~\ref{Subsect:Gap:finitedifference}. 
The gap defined with (\ref{eq:GapShift}) serves than as a point of 
reference, as it can be calculated directly within the mean--field model 
\begin{equation} 
\label{eq:block:gap} 
\gapb (Z,N) 
:= E_{\rm block} (Z,N) - E_0 (Z,N) 
\end{equation} 
as the difference in binding energy between the one blocked state of an
odd--mass number nucleus with the largest binding energy and its 
fully paired (fictitious) BCS vacuum.
Both calculations have, of course, to be performed self--consistently.  
We will call $\gapb$ the ``blocking gap'' in the following. 
% 
%----------------------------------------------------------------------
% 
\subsection{Finite--Difference Formulae} 
\label{Subsect:Gap:finitedifference} 
The gap from Eq.~(\ref{eq:block:gap}) is a purely theoretical 
construct. The problem is that $E_0$ is not measurable  
for odd--mass nuclei. We need a definition which is also experimentally 
accessible. It should fulfill two requirements which are useful for 
the phenomenological adjustment of the pairing strength: first, it 
should be easy to calculate both theoretically and from experimental 
data, and second, it should be influenced as little as possible by the 
properties of the underlying mean field in order to decouple mean--field
and pairing properties. The odd--even staggering is related to 
differences of nuclear masses and as such easy to measure as well
as to compute, As outlined above, the odd--even staggering of 
experimental masses is not a pure measure of pairing correlations, but
also has non--negligible contributions from the 
response of the underlying mean field to the blocking of a 
single--particle state.  Here we can take advantage of the fact 
that various difference formulae are conceivable and take the recipe 
which best decouples mean--field and pairing properties.

The odd--even staggering needs to be deduced from energy systematics.
To that end, there are several finite--difference formulae in the
literature to calculate $\Delta_q$ from binding energies of adjacent 
nuclei. All available finite--difference formulae for $\Delta_q$
are derived from the Taylor expansion of the nuclear mass in
nucleon--number differences \cite{Mad88a,Jen84a}
\begin{eqnarray}
\label{eq:E:expansion}
E (N)
& = &  \sum_{n=0}^\infty \frac{1}{n!}
       \frac{\partial^n E_0}{\partial N^n} \bigg|_{N_{0}} \;
       (N - N_{0})^n + D(N_{0})
\end{eqnarray}
where $E_0$ is defined in (\ref{eq:GapShift}) and the Gap $D$
given by
\begin{equation}
D
= \left\{
  \begin{array}{cl}
   0                    & \mbox{even proton and neutron number}, \\
   \Delta_{\rm n}       & \mbox{odd  neutron number},  \\
   \Delta_{\rm p}       & \mbox{odd  proton  number}. \\
  \end{array}
  \right.
\end{equation}
The number of the other kind of nucleons is assumed to be even and the
same for all terms. Combining the expansion (\ref{eq:E:expansion}) of
several adjacent nuclei leads to energy--difference formulae which can
be used to approximate the gap $\gapb$. The two--point (first--order)
formula leads to the one--nucleon separation energy, which mixes
mean--field, single--particle and pairing effects strongly and should
better not be used to fit the pairing strength. The next higher--order
is the three--point difference
\begin{eqnarray} 
\label{eq:Diff3} 
\lefteqn{E(N_0+1) - 2 E(N_0) + E(N_0 -1)}  
      \nonumber \\ 
& = &  \frac{\partial^2 E_0}{\partial N^2} \bigg|_{N_0}  
     + \frac{1}{12} \; \frac{\partial^4 E_0}{\partial N^4} \bigg|_{N_0} 
       \\ 
&   & \quad   
     + \cdots 
     + D(N_0+1) - 2 D(N_0) + D(N_0-1) 
     \quad , \nonumber  
\end{eqnarray} 
Assuming that the gap $D$ varies only slowly with nucleon number  
and that the remaining contribution from the second derivative of $E_0$ 
is negligible (which is not so well fulfilled in some cases, see
Sect.~\ref{Subsect:Res:fdf} and \cite{Sat98a,Sat98b}) this equation  
can be resolved into an approximative expression for the pairing gap 
\begin{eqnarray} 
\label{eq:3punktGap} 
\gapthree_q (N_0) 
& := & \frac{\pi_{N_0}^{\mbox{}}}{2} \Big[ E(N_0-1) 
       - 2 E(N_0) + E(N_0+1) \Big]
\end{eqnarray} 
where \mbox{$\pi_{N_0} = (-1)^{N_0}$} is the number parity. 
$\gapthree$ calculated from pure HF states without pairing 
but considering polarization of the mean field is discussed 
in Refs.\ \cite{Sat98a,Sat98b} in great detail.

The next order corresponds to a four--point difference formula, but 
this order, employing an even number of nuclei, gives an expression 
which is asymmetric around the nucleus with $N_0$ and therefore offers 
two choices. With the same assumptions used going from (\ref{eq:Diff3}) 
to (\ref{eq:3punktGap}), one possibility for the four--point gap is 
\begin{eqnarray} 
\label{eq:4punktGap} 
\gapfour_q (N_0) 
& := & \frac{\pi_{N_0}}{4} \Big[ E(N_0-2) - 3 E(N_0-1)  
       \nonumber \\  
&    & \quad  
       + 3 E(N_0) - E(N_0+1) \Big] 
       \quad . 
\end{eqnarray} 
This is an approximation for the gap at \mbox{$N_0-1/2$}. The other  
possible four--point formula gives the gap at \mbox{$N_0+1/2$}. The 
lowest--order derivative of $E_0$ hidden in the four--point formula 
is now of third order.

Although this four--point definition is widely used in the literature 
\cite{Kri90a,Cwi96a,BM} we prefer the five--point formula 
\begin{eqnarray}
\label{eq:5punktGap}
\gapfive_q (N_0)
& := & - \frac{\pi_{N_0}}{8}
       \Big[ E(N_0+2) - 4 E(N_0+1) + 6 E(N_0)
       \nonumber \\
&    & \quad
             - 4 E(N_0-1) + E(N_0-2)
       \Big]
\quad ,
\end{eqnarray}
which is symmetric and yields the best decoupling from mean--field 
effects as we will see. The smooth contributions from the mean field 
to the gap are further suppressed compared to the lower--order
formulae, the remaining derivative of $E_0$ entering $\gapfive$
is of fourth order.
% 
%---------------------------------------------------------------------------- 
% 
\subsection{Quasiparticle Energies} 
\label{Subsect:OQPE} 
Another widely used approximation for the pairing gap  
is to calculate the energy difference (\ref{eq:block:gap}) 
by constructing the blocked many--body wave function of an  
odd--mass number nucleus non--self--consistently from its BCS ground  
state as a so--called single--quasiparticle excitation 
\cite{EGIII,Ringbook,Nilssonbook}. The lowest single--quasiparticle 
energy  
\begin{equation} 
\Equasi 
= \mbox{min} ( E_k ) 
\end{equation} 
-- which we will simply denote as ``quasiparticle energy'' $\Equasi$ 
in the following -- is then another approximation for the odd--even  
staggering. In the LN scheme the $E_k$ are given in first--order  
approximation by 
\begin{equation} 
\label{eq:EquasiLN} 
E_k 
\approx \sqrt{(\epsilon_k^\prime - \lambda_q)^2 + f_k^2 \Delta_{k}^2} 
        + \lambda_{2,q} 
\quad , 
\end{equation} 
see Appendix~\ref{Sect:OQPE} for details. The important difference 
between $\Equasi$ and $\gapb$ is that for $\Equasi$ the blocked 
many--body  wave function is not calculated self--consistently. 
%
%------------------------------------------------------------------- 
% 
\subsection{Spectral Gaps} 
\label{Subsect:Gap:spectral} 
The calculation of $\gapfive$ from mean--field models is a bit 
cumbersome since it requires information on five nuclei including 
nuclei with odd mass number. Moreover, the definition becomes 
inapplicable to describe the variation of pairing correlations with 
deformation for a given nucleus. At this point, we could recur to the 
purely theoretical definition (\ref{eq:block:gap}) involving a blocked 
and an unblocked BCS calculation. This still requires involved 
calculations and can become unwieldy in deformed calculations.

Therefore, a commonly used approach is to estimate the pairing gap  
from spectral properties of a nucleus. In schematic pairing models  
using the same pairing matrix element for all states the  
single--particle gaps (\ref{eq:spGap}) turn out to be state--independent  
and are thus immediately a measure for the pairing correlations.  
With local forces as used here we obtain state--dependent  
single--particle gaps $\Delta_k$ and have to define 
an average gap as representative for the strength of the pairing  
correlations. The authors of \cite{Dob84a} have proposed to use the  
average of the single--particle gaps (\ref{eq:spGap}) weighted with 
the occupation probability $v^2_k$ 
\begin{equation} 
\label{eq:v2Gap} 
\gapvv_q 
= \frac{\sum_{k \in \Omega_q} f_k \, v_k^2 \, \Delta_k} 
       {\sum_{k \in \Omega_q} f_k \, v_k^2} 
\quad . 
\end{equation} 
This definition, however, puts too much weight on deeply--bound states 
whereas pairing is a mechanism most active near the Fermi surface. 
We therefore propose an average with the same factor $v_k u_k$ as 
it appears in the accumulation of the pair density $\chi (\textbf{r})$,  
see Eq.~(\ref{eq:PairDensity}), yielding the spectral gap as 
\begin{equation} 
\label{eq:uvGap} 
\gapuv_q 
= \frac{\sum_{k \in \Omega_q} f_k \, v_k u_k \, \Delta_k} 
       {\sum_{k \in \Omega_q} f_k \, u_k v_k} 
\quad . 
\end{equation} 
Note that the spectral gaps are an estimate for the pairing gap and
therefore the contribution of pairing correlations to the odd--even
staggering (\ref{eq:block:gap}). 
Assuming that the spectral gaps (\ref{eq:v2Gap}) and (\ref{eq:uvGap}) 
are an approximation for the square--root term in (\ref{eq:EquasiLN}), 
approximately particle--number projected spectral gaps are given by 
\begin{eqnarray} 
\label{vvGapLN} 
\gapvv_{q}^{\rm (LN)} 
= \gapvv_q + \lambda_{2,q} 
  \quad , \\ 
\label{uvGapLN} 
\gapuv_{q}^{\rm (LN)} 
= \gapuv_q + \lambda_{2,q} 
  \quad , 
\end{eqnarray} 
which was proposed by the authors of \cite{Cwi96a} for the average  
gap $\gapvv^{\rm (LN)}$. These spectral gaps will be discussed and  
compared with other alternatives for the calculated pairing gap in  
Section~\ref{sect:results}. 
%
%======================================================================= 
% 
\section{Results and Discussion} 
\label{sect:results} 
\subsection{Fit of Pairing Strength} 
\label{subsect:pairfit} 
% 
% 
%-------------- 
% 
\begin{table}[t!] 
\caption{ 
\label{tab:strength} 
Pairing strengths $V_{\rm n}$ for the neutrons and $V_{\rm p}$ for the 
protons in \mbox{${\rm MeV} \; {\rm fm}^3$} for the DF and DDDI pairing 
functionals used in the BCS and LN schemes in connection with SkI4. 
} 
\centerline{ 
\begin{tabular}{lcccc}  
\hline\noalign{\smallskip} 
                 & \multicolumn{2}{c}{DF}  
                 & \multicolumn{2}{c}{DDDI} \\  
\noalign{\smallskip}\hline\noalign{\smallskip} 
Scheme           & $V_{\rm n}$ 
                 & $V_{\rm p}$  
                 & $V_{\rm n}$ 
                 & $V_{\rm p}$  
\\ \noalign{\smallskip}\hline\noalign{\smallskip} 
BCS              & $-323$ & $-310$ & $-999$ & $-1146$ \\ 
LN               & $-318$ & $-250$ & $-947$ & $- 947$ \\ 
\noalign{\smallskip} \hline 
\end{tabular} 
}
\end{table} 
% 
%-------------- 
% 
The first step is to determine appropriate pairing strengths $V_{0,q}$ 
for the DF and the DDDI functionals. We do that on the grounds of the 
five--point gap $\gapfive$ and adjust the pairing strength by fitting 
calculated values for $\gapfive$ to experimental ones for a large set of 
semi--magic nuclei, i.e.\ the isotope chains 
${}^{44}_{22}\mbox{Ca}$,  
${}^{106}_{\; \: 56}\mbox{Sn}$--${}^{128}_{\; \: 78}\mbox{Sn}$, and 
${}^{201}_{119}\mbox{Pb}$--${}^{206}_{124}\mbox{Pb}$,  
for the neutrons and the isotone chains 
${}^{52}_{28}\mbox{Cr}$,  
${}^{82}_{50}\mbox{Ge}_{32}$--${}^{94}_{50}\mbox{Ru}_{44}$, 
${}^{136}_{\; \: 82}\mbox{Xe}_{54}$--${}^{147}_{\; \: 82}\mbox{Tb}_{65}$, and  
${}^{210}_{126}\mbox{Po}_{84}$--${}^{215}_{126}\mbox{Ac}_{89}$ 
for the protons. 
A seperate fit has been performed for each one of the pairing 
functionals, DF or DDDI, and for each appraoch, BCS or LN.
Each nucleon sort, proton or neutron, aquires its own pairing
strength adjusted to isotonic chains, or isotopic chains respectively.

The experimental data are taken from \cite{Wapstra}. The resulting  
values for the pairing strength are listed in Table~\ref{tab:strength}.
We obtain a reasonable fit of the pairing gaps for all pairing schemes, 
see also Figures~\ref{fig:bcs} (compare ``Expt.´´ with $\gapfive$)
and~\ref{fig:vergl} in what follows. We have checked that the pairing 
strengths are not significantly changed (i.e.\ on the order of $1 \%$)  
when fitting instead theoretical $\gapthree$ to experimental
$\gapthree$ or similarly the $\gapfour$.
% 
%----------------------------------------------------------------------- 
% 
\subsection{Comparison of the Finite--Difference Formulae} 
\label{Subsect:Res:fdf}
%
% 
%-------- 
% 
\begin{figure*}[ht] 
\centerline{\epsfig{file=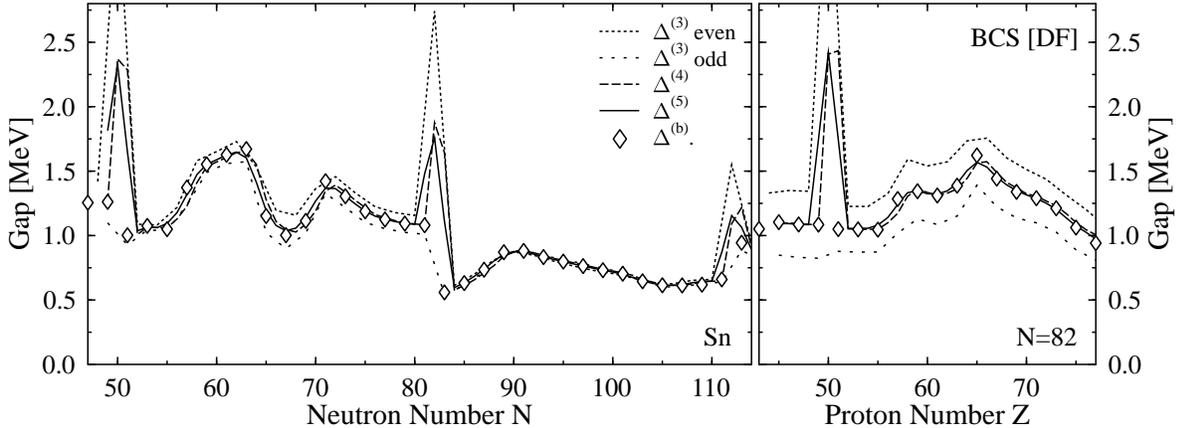}} 
\caption{\label{fig:delta:vergl} 
Comparison of the three--point $\gapthree$, four--point $\gapfour$  
and five--point difference formulae $\gapfive$ and the blocking gap  
$\gapb$. The left panel shows results for neutron gaps in the  
chain of tin isotopes, the right panel for proton gaps in the chain of the 
\mbox{$N=82$} isotones.} 
\end{figure*} 
% 
%-------- 
% 
The various finite--difference gaps were introduced as 
experimentally accessible approximations
to $\gapb$, i.e.\ the energy differences of blocked and unblocked 
calculations of odd--mass nuclei. Figure~\ref{fig:delta:vergl} 
compares directly the performance of the $\Delta^{(i)}$, \mbox{$i=3,4,5$}, 
in this respect for calculations within the BCS approach. 
The three--point gaps $\gapthree$ show a pronounced odd--even staggering
for all \mbox{$N=82$} isotones and the tin isotopes with neutron 
numbers below the \mbox{$N=82$} shell closure. Note that we have 
disentangled that by drawing a separate line for even--even and 
odd--mass nuclei. These two lines for $\gapthree$ embrace the gap 
$\gapb$. The staggering is caused by non--vanishing mean--field 
contributions to $\gapthree$ (mainly the second derivative term
in Eq.~(\ref{eq:Diff3})), which enter with a different sign for 
even--even and odd--mass nuclei. The four--point gaps $\gapfour$ from 
Eq.~(\ref{eq:4punktGap}) give a smoother approximation for $\gapb$. 
But as expected the 
$\gapfour$ are slightly shifted versus the $\gapb$ which becomes 
rather obvious where the gaps change rapidly, e.g.\ around  
\mbox{$N = 63$}, \mbox{$N = 70$}, \mbox{$Z = 57$}, \mbox{$Z = 65$}. 
The five--point gaps $\gapfive$ give the best overall agreement with 
the $\gapb$. 
% 
%------- 
% 
\begin{figure}[b!] 
\centerline{\epsfig{file=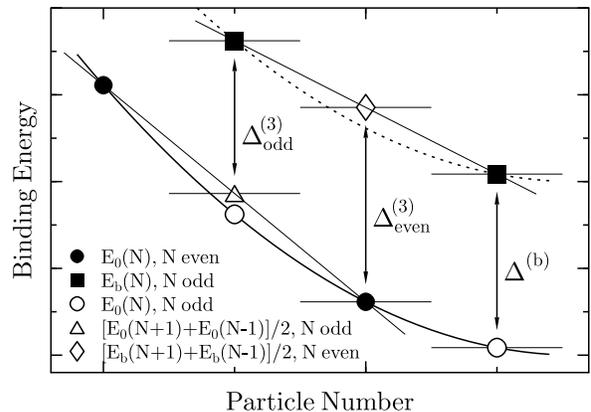}} 
\caption{\label{fig:delta3}
Schematic comparison of $\gapthree$ and $\gapb$.
$E_0$ and $E_{\rm b}$ denote BCS ground--state energies and the energy
of the blocked ground state of an odd--mass nucleus respectively.
Filled symbols denote binding energies that are experimentally accessible,
while the BCS ground state of an odd--mass nucleus (white circles) 
can be calculated from the mean--field model only.
}
\end{figure} 
% 
%------- 
%

The oscillation of the $\gapthree$ around the values for $\gapb$ 
has a simple geometrical reason, as can be seen from Fig.~\ref{fig:delta3}.
$\gapb$ is per definition (\ref{eq:block:gap}) the shift between
the smooth curve connecting the (unblocked)
BCS ground--state energies $E_0$ 
of all nuclei (thick line through circles)
and the smooth curve that connects the (blocked) ground--state 
energies $E_{\rm b}$ of odd--mass nuclei
(dotted line through full boxes). The three--point gap of 
an even--even nucleus $\Delta^{(3)}_{\rm even} (N_0) 
= \tfrac{1}{2} [E_{\rm b}(N_0-1)+E_{\rm b}(N_0+1)] - E_0(N_0)$
is the shift between the smooth curve connecting the $E_0$ and the
average energy of the two adjacent odd--mass nuclei, which lies
outside the band given by $\gapb$. From this follows immediately
$\Delta^{(3)}_{\rm even} > \gapb$ for bound nuclei. A similar
construction leads to $\Delta^{(3)}_{\rm odd} < \gapb$, see
Fig.~\ref{fig:delta3}. For $\gapfour$ and $\gapfive$ the deviation 
from $\gapb$ becomes of course much smaller because the higher--order 
finite--difference formulae give a better approximation of the 
smooth curves connecting the  $E_0$ and $E_{\rm b}$ respectively. 
This qualitative result does not depend on the level
of sophistication for the calculation of the 
odd--mass--number nuclei. Considering polarization effects will 
shift $E_{\rm b}$ with respect to $E_0$ (which has to be 
counterweighted by a refit of the pairing strength 
\cite{Xu99a,uggap}) and may distort the surface of the 
$E_{\rm b}$, but will not change the sign of its curvature.

There remains a significant difference between all
$\Delta^{(i)}$, $i=3,4,5$ and $\gapb$  around shell 
closures where finite--difference formulae (except 
$\gapthree$ for odd nuclei) produce a peak which becomes 
broader with increasing order of the difference formula.
We want to discuss the origin of this peak for the example of the 
five--point gap $\gapfive$. The five--point approximation 
(\ref{eq:5punktGap}) for $\gapb$ makes three assumptions: 
(i) the binding energy is an 
analytical function of the nucleon numbers and therefore can be 
expanded in a Taylor series in the range of two mass units around the 
considered nucleus; (ii) derivatives of $E_0$ of higher than third order 
are negligible; and (iii) the gap varies only slowly with nucleon 
number. The first two assumptions are strongly violated at shell 
closures, where the kink in the systematics of binding energies 
does not allow the Taylor expansion (\ref{eq:E:expansion}) and
leads to a spurious contribution from the mean--field functional
to the finite--difference gaps.
% 
%------- 
%
\begin{figure}[ht]
\centerline{\epsfig{file=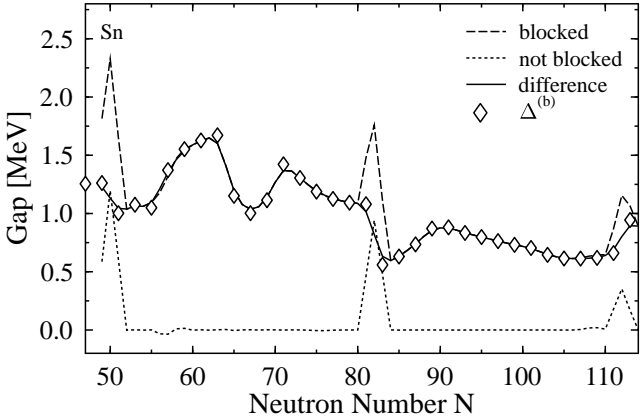}}
\caption{\label{fig:gap:noblock}
Five--point gaps $\gapfive$ computed using either blocked calculations
of the odd--mass nuclei (dashed lines) or their BCS ground states (dotted
lines) for the chain of tin isotopes with BCS+DF pairing.
Non--zero $\gapfive$ from not blocked calculations appear only at
closed shells and are caused by a spurious contribution of the mean
field to $\gapfive$. The difference of both curves (solid line) gives
the contribution from the blocking to $\gapfive$.
}
\end{figure}
% 
%------- 
% 

To visualize this effect we compare in Fig.~\ref{fig:gap:noblock} 
$\gapfive$ with five--point gaps $\gapfive{}^{\rm (nb)}$  
using the binding energies of not blocked calculations of the 
odd--mass--number nuclei, which carries only the the spurious mean--field
contribution to $\gapfive$. Subtracting $\gapfive{}^{\rm (nb)}$ 
from $\gapfive$ one gets the contribution from the blocking 
of the odd particle to $\gapfive$. 
The peaks at closed shells disappear yielding a smooth curve for  
$\gapfive-\gapfive{}^{\rm (nb)}$ which now follows the values of  
$\gapb$ everywhere.

Besides the immediate vicinity of shell closures, we find that 
$\gapfive$ is a very good approximation for the staggering gap $\gapb$.   
For a few additional nuclei there remains a small difference between 
$\gapfive$ and $\gapb$, see Fig.~\ref{fig:gap:noblock}. This occurs 
for example around \mbox{$N=68$} and \mbox{$N=72$} for the tin  
isotopes. For those nuclei the assumption of a slow variation of 
the gap with nucleon number which enters the five--point 
formula (\ref{eq:5punktGap}) is not valid: the gap changes in these 
regions by about $40 \%$ (due to a sudden change of the density 
of single--particle levels at the Fermi surface in these nuclei). But 
even then the deviation between $\gapfive$ and $\gapb$ remains 
acceptably small. We thus prefer $\gapfive$ for the fit of pairing
strengths. The small difference between $\gapfive$ and $\gapb$
even  suggests a simplified fitting procedure where calculated 
$\gapb$ are adjusted to experimental values for $\gapfive$ in 
chains of spherical semi--magic nuclei.

As explained in Sect.~\ref{Subsect:gapb}, the $\gapb$ unavoidably
contain a contribution from the polarization of the mean field in
odd nuclei. It is therefore somewhat unlucky and confusing that
$\gapb$ is usually denoted as ``pairing gap'' in the literature,
see e.g.\ \cite{Mol92a,Mad88a,BM}. The commonly used fist formulae
like $\gapb \approx 12/\sqrt{A}$ are intended to represent the average
trend of the odd--even staggering, not the pairing gap. The discrepancies
between the two quantities can be expected to decrease with increasing
mass number and to be very small for heavy nuclei.

An analysis of gaps from a complementary point of view is 
given in \cite{Sat98a} for the case of $\gapthree$. 
This study omits pairing altogether and concentrates 
exclusively on polarization effects for $\gapthree$ using pure
deformed HF calculations. The focus is on small nuclei because these
have most pronounced deformation effects. In this framework, 
it is found that 3--point gaps show a pronounced 
odd--even staggering with the $\Delta^{(3)}_{\rm odd}$ being close 
to zero while most of the $\Delta^{(3)}_{\rm even}$ have large positive 
values in most cases. This finding is explained in terms 
of the macroscopic--microscopic model in that
a large contribution from the symmetry energy
is counterweighted by the difference of single--particle 
energies when calculating $\Delta^{(3)}_{\rm odd}$. This 
observation led the authors of \cite{Sat98a} to the conclusions that 
$\Delta^{(3)}_{\rm odd}$ is very close to the pure pairing gap
while $\Delta^{(3)}_{\rm even }$ contains a contribution 
from the mean field. The higher--order gaps $\gapfour$
and $\gapfive$ turn out to be rather useless in that environment.
We want to point out that
the odd--even staggering of $\gapthree$ observed in \cite{Sat98a}
is not the staggering of $\gapthree$ around the values for $\gapb$
as we discuss it here, see e.g.\ Fig.~\ref{fig:delta:vergl}.
The deformation staggering is a
phenomenon much similar to the spurious peak of finite--difference
gaps at major shell closures discussed above.
Mind that pure deformed HF calculations in small nuclei
produce a subshell closure for each even--even nucleus by virtue
of the Jahn--Teller effect.
This, in turn, leads to a spurious contribution 
to $\Delta^{(3)}_{\rm even}$ for nearly all even--even nuclei which
is half of the energy difference between single--particle levels 
\cite{Sat98a}, and it has devastating consequences for the systematics 
of 4--point and 5--point gaps.
The picture changes dramatically when pairing is included, as we do
here. Pairing induces a drive to spherical shapes and thus reduces
deformation effects dramatically while rendering blocking the
the dominant contribution to the gaps. This smoothes 
the systematics of binding energies and thus of any $\Delta^{(i)}$.
Moreover, we are discussing here semi--magic medium and heavy nuclei
which have spherical BCS ground states. We are thus
considering a sample with minimal mean--field effects, just appropriate
to concentrate on pairing features.
The interplay of polarization effects which we neglect here 
and pairing correlations will probably play a role for small
non--magic nuclei. It deserves further inspection in the future.
%
%-----------------------------------------------------------------------
%
\subsection{Spectral Gaps in BCS Pairing}
% 
%-------------- 
% 
\begin{figure*}[t!] 
\centerline{\epsfig{file=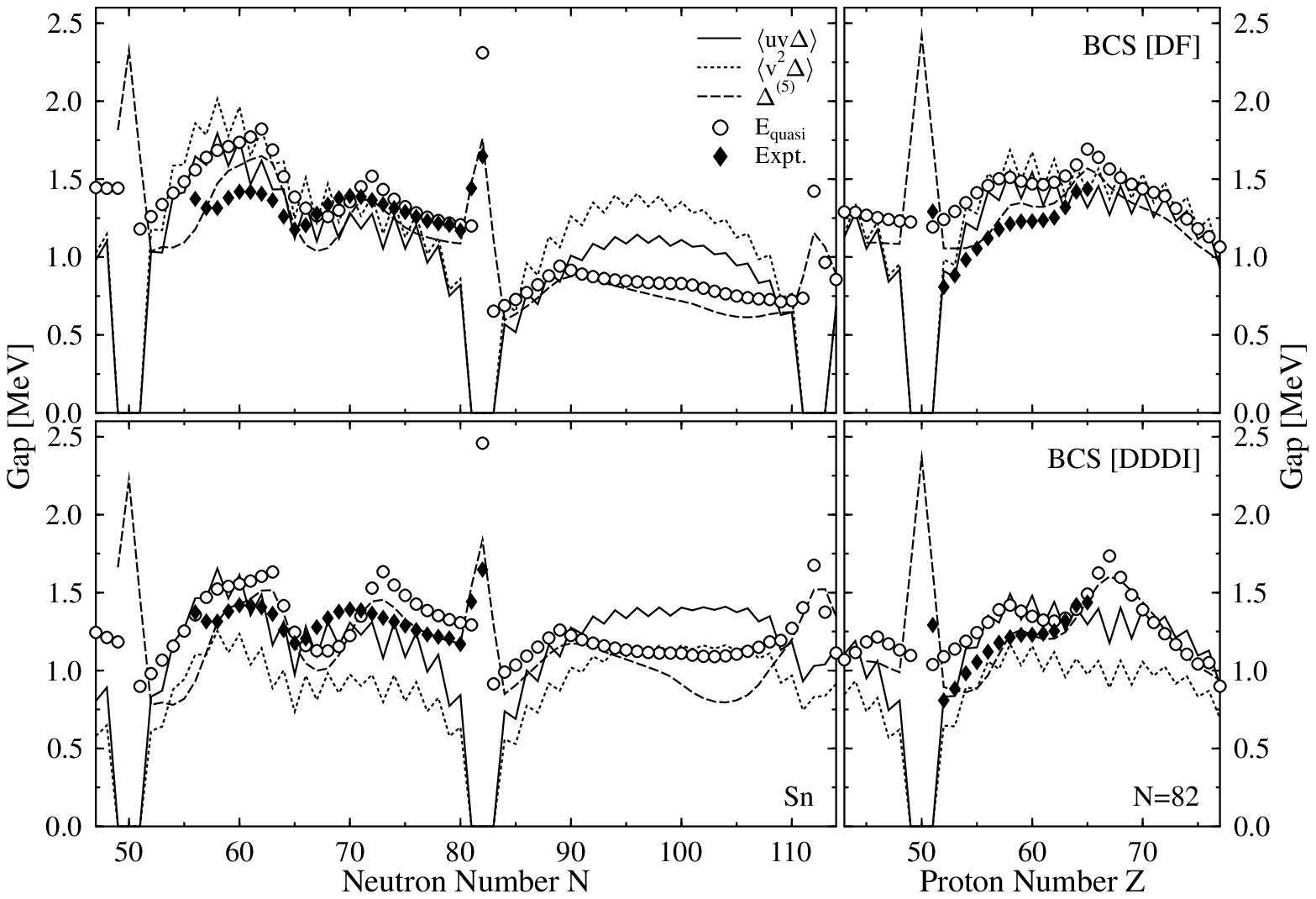}} 
\caption{\label{fig:bcs} 
Comparison of the five--point gap $\gapfive$, the spectral gap $\gapuv$,  
the average gap $\gapvv$, and the single--quasiparticle energy 
$\Equasi$ with experimental values for the five--point gap 
for tin isotopes (left panels) and \mbox{$N=82$} isotones (right 
panels) calculated with a delta pairing force (upper panels) and DDDI 
force (lower panels) in the BCS scheme. The experimental values are 
calculated from the binding energies given in \protect\cite{Wapstra} 
using the five--point formula $\gapfive$ (\protect\ref{eq:5punktGap}).
} 
\end{figure*} 
% 
%-------------- 
% 
Having discussed the various finite difference gaps, we now take the 
five--point gap as reference value and study the relation to the 
spectral gaps and quasiparticle energies. Figure~\ref{fig:bcs} 
compares calculated results for $\gapfive$, $\gapvv$, $\gapuv$ and 
$E_{\rm quasi}$ computed within the BCS approach.

Let us start the discussion by looking at the neutron gaps in the 
chain of tin isotopes calculated with the DF pairing interaction
in the BCS approach (upper left panel).
The spectral gaps $\gapuv$ and $\gapvv$ show a pronounced 
odd--even staggering where the gaps of even--even nuclei have larger 
values than those of odd--mass nuclei.  This is caused by the 
blocking of one state with a large weight $uv$ in the odd--mass 
nucleus. The blocked state does not contribute to the pairing 
potential, leading to overall smaller single--particle gaps 
(\ref{eq:spGap}). The amplitude of the odd--even staggering decreases, 
of course, with increasing neutron number $N$ because the relative 
contribution of a particular state to the pair density becomes smaller 
with increasing level density in the heavier isotopes.

Both $\gapvv$ and $\gapuv$ are exactly zero for closed shell nuclei,  
i.e.\ \mbox{$N = 50$}, \mbox{$N = 82$}, \mbox{$N = 112$}, 
and the adjacent odd mass--number nuclei. In these nuclei the BCS  
scheme breaks down. This is a deficiency of the BCS scheme which 
is related to the particle--number uncertainty of the BCS state  
\cite{Ringbook}.

The quasiparticle energies $\Equasi$ show a similar dependence on the 
neutron number as the $\gapfive$, but for most nuclei they are 
100--250 keV larger (and thus the same amount larger with respect to 
the $\gapb$). This is caused by calculating the blocked many--body 
wave function entering $\Equasi$ not self--consistently. 
The variational principle behind the self--consistent calculation of 
the odd--mass nuclei entering the $\gapfive$ leads always to larger a 
binding energy of the odd--mass nuclei, lowering the calculated gap.

The quasiparticle energies show the same peak  
at shell closures as the finite--difference gaps which is  
again related to a spurious contribution from the mean field. 
For non--magic nuclei the lowest single--quasiparticle state 
usually corresponds to a single--particle state with  
\mbox{$\epsilon_k^\prime \approx \lambda_q$} leading to 
the single--quasiparticle energy  
\mbox{$\Equasi \approx \Delta_{k} + \lambda_{2,q}$}. In 
magic nuclei one has an additional contribution from the first term in 
the square root in Eq.~(\ref{eq:EquasiLN}) 
since the Fermi energy is approximately in the middle of the  
gap in the single--particle spectrum (In the BCS scheme, where the 
pairing breaks down for closed--shell nuclei the derivation of 
Eq.~(\ref{eq:EquasiLN}) is not valid. Then one has different 
Fermi energies for the removal and addition of a particle, which 
are the single--particle energies of the last occupied and first 
unoccupied single--particle state respectively). The large jump in  
the Fermi energy is the reason for the kink in the systematics  
of binding energies at shell closures, which in turn causes the peak  
in the finite--difference gaps discussed above.

While all definitions of the gap give similar values in  
the valley of stability, there appear large differences between  
the $\gapfive$ and $\Equasi$ on one hand and the $\gapvv$  
and $\gapuv$ on the other 
hand for neutron--rich nuclei beyond the \hbox{$N = 82$} shell closure.  
The spectral gaps overestimate the $\gapfive$ (which closely follow the 
$\gapb$ as discussed earlier). The $\gapuv$ are smaller than the $\gapvv$  
for all tin isotopes and in most systems the $\gapuv$ are closer to the  
$\gapfive$ than the $\gapvv$.

\begin{figure}[b!] 
\centerline{\epsfig{file=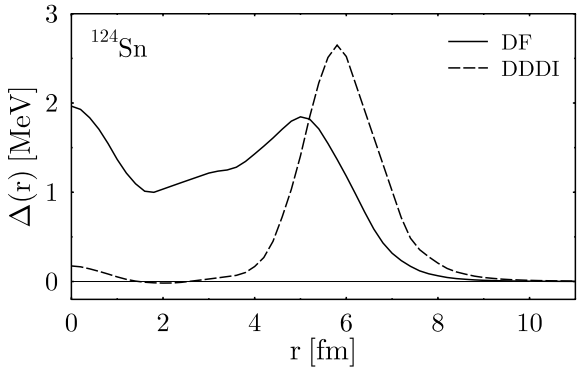}} 
\caption{\label{fig:pairpot} 
Local pairing potential $\Delta (r)$ of the neutrons in ${}^{124}{\rm Sn}$ 
for the DF and DDDI pairing functionals. The DF pairing potential acts 
over the whole volume of the nucleus, while the DDDI pairing potential 
is peaked at the nuclear surface. 
} 
\end{figure}

\begin{figure}[ht] 
\centerline{\epsfig{file=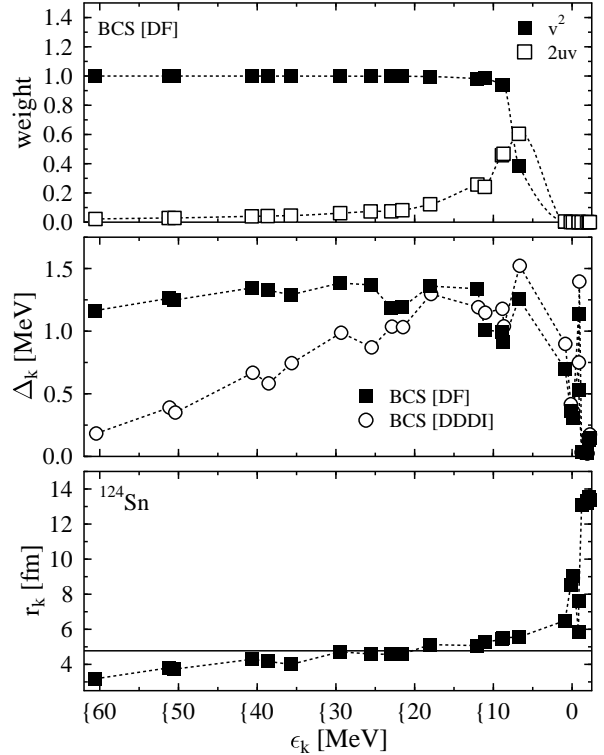}} 
\caption{\label{fig:spgap} 
Weights $v^2$ (full diamonds) and $2uv$ (open diamonds) entering  
the calculation of the spectral gaps 
$\gapvv$ and $\gapuv$ (upper panel), single--particle gap $\Delta_k$ 
in MeV (middle panel) and single--particle root--mean--square radius 
(lower panels) of the neutrons in ${}^{124}{\rm Sn}$, calculated with  
SkI4 in the BCS scheme. Weights and radii are shown for results with  
the DF functional only (DDDI gives very similar results),  
while the $\Delta_k$ are shown for 
the DF (diamonds) and DDDI (circles) pairing functionals.  
The horizontal line in the lower panel is the root--mean--square  
radius of the neutron density. 
The $uv$ in the upper panel are multiplied by two to have the  
same possible maximum value of one as the $v^2$.
} 
\end{figure}

Now we want to look at the changes when employing the DDDI pairing 
functional instead of the simpler delta force, see the lower left 
panel of Fig.~\ref{fig:bcs}. There are two significant differences to 
the results obtained with the DF functional: (i) all gaps except 
$\gapvv$ are larger by roughly $20 \, \%$ for neutron--rich nuclei 
with \mbox{$N > 82$}, and (ii) the $\gapvv$ are always much smaller than 
the $\gapuv$. This trend is just the opposite from the one for the DF 
functional.
This is caused by the different choice of weights in the  
definition of $\gapvv$ and $\gapuv$ in combination with the 
spectral distribution of the single--particle gaps $\Delta_k$. 
While the DF pairing potential follows roughly the nuclear 
density distribution, the DDDI functional gives a pairing  
potential which is sharply peaked at the nuclear surface, 
see Fig.~\ref{fig:pairpot}. This leads in case of the DF interaction 
to single--particle gaps $\Delta_k$ of comparable size for all  
bound states while in case of the DDDI interaction the $\Delta_k$ 
of deeply bound single--particle states are rather small, see the middle 
panel of Fig.~\ref{fig:spgap}. Together with the weight factors 
used to calculate $\gapuv$ and $\gapvv$ -- see the upper panel 
in  Fig.~\ref{fig:spgap} -- this gives the observed pattern 
for the spectral gaps: when calculated with the DF interaction  
they are rather insensitive to the choice for the weight factors  
while there is a huge difference in case of the DDDI interaction.
% 
%--------- 
% 
% 
%--------- 
% 

This explains also why the $\gapfive$ extrapolate 
quite differently when comparing DF and DDDI pairing for neutron--rich 
nuclei. In these nuclei the states at the Fermi surface are only 
loosely bound and therefore have a large spatial extension but only 
small overlap with the volume--like DF pairing potential. An 
extreme example is the tin isotope with \mbox{$N=112$} where 
DF pairing breaks down but DDDI pairing is still fully active. 
Experimental data on the excitation spectra of neutron--rich nuclei  
will give in the future valuable information to distinguish between  
volume--like (DF) and surface--peaked pairing interactions. 
% 
%--------- 
% 
% 
%--------- 
% 

This different behavior of $\gapvv$ on one hand and $\gapuv$ and 
$\gapfive$ on the other hand hints that is possibly dangerous to use 
different definitions of the gap for experimental and 
calculated gaps when fitting the pairing strength and comparing 
calculated and experimental values.

The right panels of Fig.~\ref{fig:bcs} show the pairing gaps of the 
protons in the chain of \hbox{$N = 82$} isotones. Qualitatively the 
results are similar to those for the neutron gaps in the tin isotopes. 
But the overall reproduction of the experimental values is much better  
than in the case of tin isotopes and the differences between the various  
gaps is much smaller when going towards the drip--line. The 
Coulomb potential stabilizes even loosely--bound protons. Therefore the  
difference between the gaps comparing volume--like and surface--like  
pairing potentials is quite small. Only for the DDDI force remains 
the large difference between the spectral gaps $\gapvv$ and $\gapuv$  
explained above. 
% 
%----------------------------------------------------------------------- 
% 
\subsection{Lipkin--Nogami Pairing} 
\label{SubSect:Results:LN} 
% 
%--------- 
% 
\begin{figure*}[t!] 
\centerline{\epsfig{file=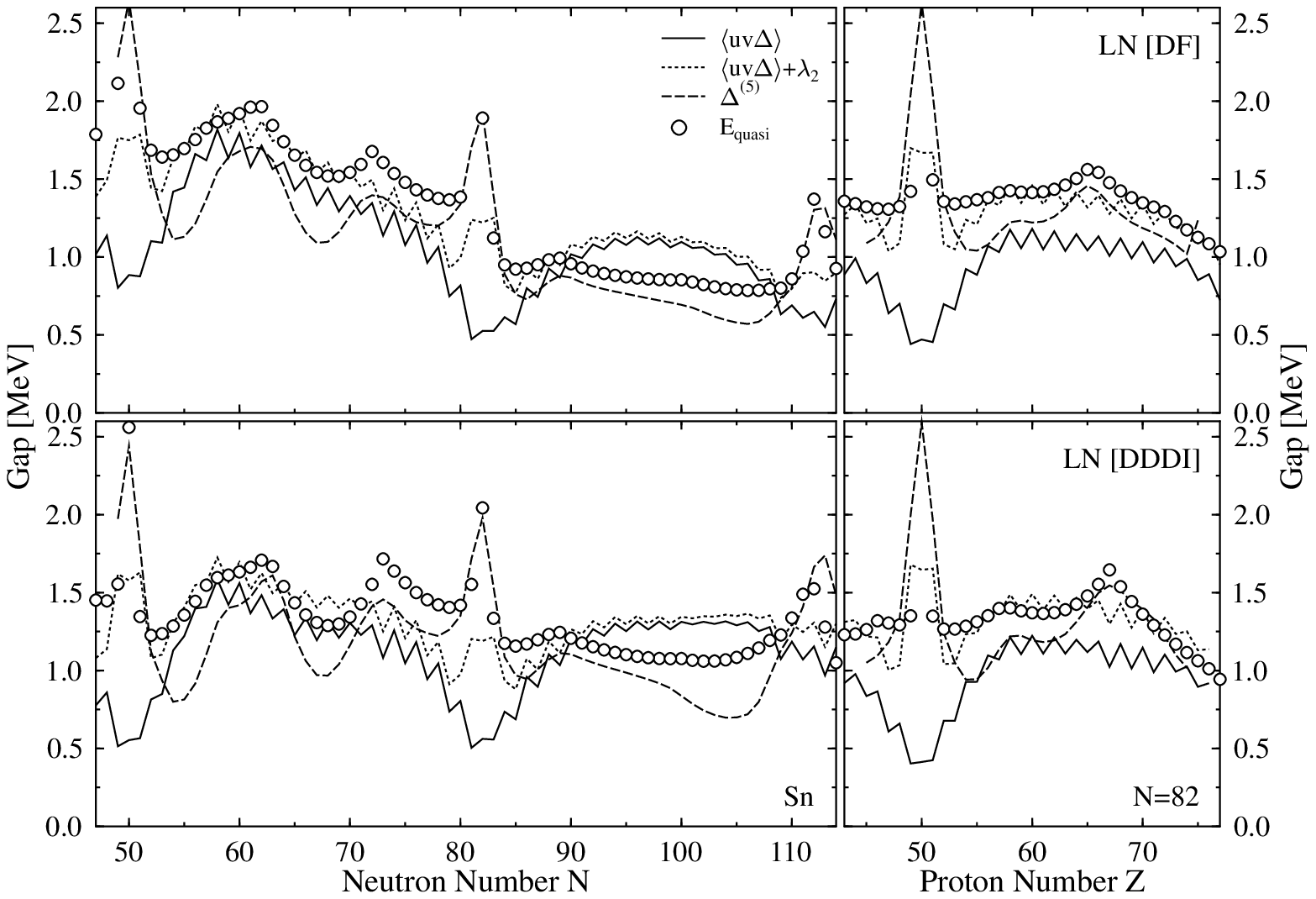}} 
\caption{\label{fig:ln} 
Comparison of the five--point gap $\gapfive$, the spectral gap 
$\gapuv$, the particle-number corrected spectral gap  
\mbox{$\gapuv+\lambda_2$} and the single--quasiparticle energy 
$\Equasi$ with experimental values for the five--point gap 
for tin isotopes (left panels) and \mbox{$N=82$} isotones (right 
panels) calculated with a delta pairing force (upper panels) 
and DDDI force (lower panels) in the LN scheme. 
} 
\end{figure*} 
% 
%--------- 
% 
Figure~\ref{fig:ln} shows the gaps of the neutrons in the 
chain of tin isotopes, now calculated with 
the LN scheme. The spectral gaps $\gapvv$ are omitted here. Instead we 
compare the ``bare'' $\gapuv$ (\ref{eq:uvGap}) with the particle--number  
corrected values \hbox{$\gapuv + \lambda_{2}$} (\ref{uvGapLN}).

The global pattern of the various gaps looks very similar to 
the one obtained with the BCS scheme, see Fig.~\ref{fig:bcs}. The most 
obvious difference between the BCS and LN methods is that the LN 
scheme does not break down for closed--shell nuclei. Therefore the 
``bare'' spectral gap $\gapuv$ has a finite -- but still somewhat too 
small -- value around magic nuclei. Adding $\lambda_2$ gives better 
results around shell closures, but away from shell closures the 
difference between $\gapuv$ and \hbox{$\gapuv + \lambda_{2}$}  
for the tin isotopes is too small to decide on 
one preferred definition.

From the difference between $\gapuv$ and \hbox{$\gapuv + \lambda_{2}$} 
it can be seen that $\lambda_2$ is largest around shell closures. But 
this indicates also that the LN approximation might not be sufficient for 
magic nuclei, a variational calculation of $\lambda_2$ or even full  
projection of the many--body wave function is needed 
\cite{Dob92a} there.

The single--quasiparticle energies $\Equasi$ follow closely the  
particle--number projected \hbox{$\gapuv_{\rm p} + \lambda_{2, \rm p}$}, 
which is easily understood remembering that $\lambda_2$ is added to 
both quantities. At shell closures, however, the single--quasi\-par\-ticle 
energies overestimate the experimental gaps. Like in the case of the 
BCS scheme the $\Equasi$ are nearly always larger than the calculated  
five--point gaps $\gapfive$. 
% 
%----------------------------------------------------------------------- 
% 
\subsection{Comparison of all Models} 
\label{Subsect:Results:Comparison} 
%
%-------- 
% 
\begin{figure*}[t!] 
\centerline{\epsfig{file=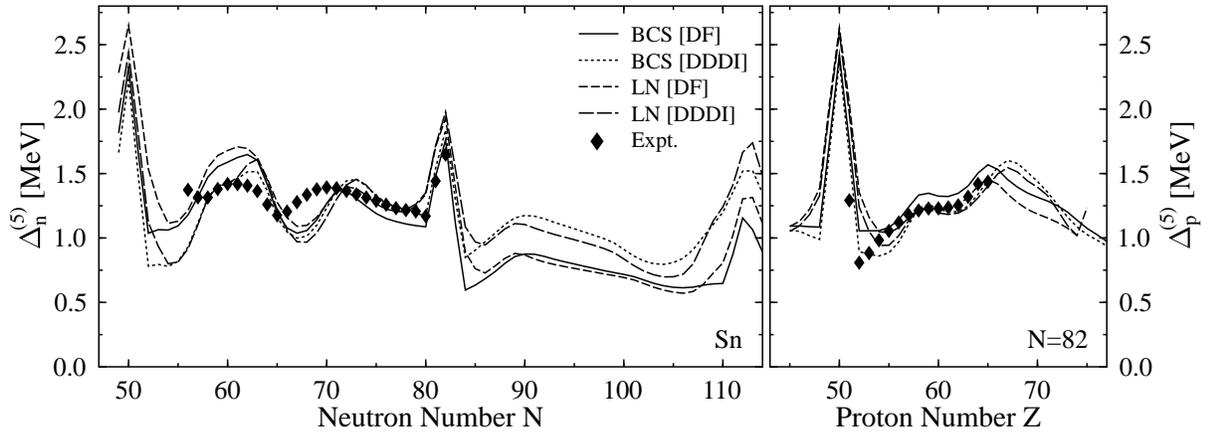}} 
\caption{\label{fig:vergl} 
Comparison of the \hbox{$\gapfive$} calculated in the 
different pairing schemes for the chains of tin isotopes
and \mbox{$N=82$} isotones. 
} 
\end{figure*} 
% 
%-------- 
%
Finally, we want to compare the four pairing models, i.e.\ any combination  
of BCS or LN and DF or DDDI pairing. The comparison is done with respect  
to the five--point gap $\gapfive$, which we prefer as the most robust  
empirical definition and which has turned out to be the most 
useful definition of the pairing gap. In Fig.~\ref{fig:vergl} 
the five--point gaps calculated from various pairing schemes 
are compared with experimental values. 
The differences between BCS and LN are generally very small.  It is to 
be remembered, however, that the effective pairing strength is 
readjusted for the LN scheme. The (approximate) particle--number 
projection increases the total binding energy, but this effect 
is renormalized by virtue of the fit delivering a slightly smaller 
pairing strength in the LN scheme. There is one detail where BCS and 
LN differ: the peak of $\gapfive$ in the vicinity of closed shells is 
more spread out in case of LN which is probably due to the softening 
of the shell closure by LN.

The differences between the pairing forces (DF versus DDDI) are much 
larger.  For the neutron gaps of tin isotopes close to the valley of 
stability and the proton gaps in the \hbox{$N = 82$} isotones all 
schemes and forces still give similar results, but large differences 
between the DF and DDDI force occur for neutron gaps around \hbox{$N = 
60$} and very neutron--rich nuclei with \hbox{$N > 82$}.  Only the 
DDDI model can describe the gaps in both the light and heavy 
known tin isotopes, while the DF interaction overestimates the gaps in 
the light ones by up to 15--20$\%$. The particle--number projection 
has only a small effect on the gaps when the strength of the pairing 
interaction is readjusted.  The better description of the tin isotopes 
around \hbox{$N = 60$} with the DDDI force gives a hint that this 
pairing interaction may be more realistic than a delta force. This, 
however, has to be taken with a grain of salt: it may also be a  
spurious effect due to a deficiency of the underlying mean--field.  
The disagreement between calculated and experimental $\gapfive$  
around \hbox{$N = 70$} is rather robust in that all forces and  
schemes give very similar results. 
For a profound decision which pairing interaction gives the 
most realistic results throughout the chart of nuclei more and other 
observables have to be investigated using various forces for the 
underlying mean--field. Research in that direction is underway.

We have checked that one obtains very similar results when comparing 
experimental and calculated three--point and four--point gaps.  
However, it is important that the experimental and calculated values 
are computed from the same formula. As already shown in  
Fig.~\ref{fig:delta:vergl}, the various finite--difference formulae 
may give results which differ by $25 \%$. 
% 
%======================================================================= 
% 
\section{Summary}  
\label{sect:conclusions} 
We have compared various approximations for the calculation of the pairing  
gap: three--point, four--point and five--point finite difference formulae  
where the gap is estimated from total binding energies of adjacent nuclei,  
spectral gaps with different weights which put bias on well--bound states  
or levels at the Fermi surface, and the single--quasiparticle energy. 
Predictions of four different pairing models for the gaps were compared, 
namely the BCS and LN pairing schemes employing the DF or DDDI interactions. 
Experimental values for the pairing gaps are usually calculated from a  
finite--difference formula. For the calculated gaps we find that apart  
from shell closures there are non--negligible deviations up to $25 \%$  
between the various definitions. Some definitions of the pairing gap cannot  
be used for closed--shell nuclei. Therefore, it is the safest choice  
to compute the pairing gaps from mean--field calculations in the same way  
as the experimental values.

The natural definition for the calculated gap is the difference  
in binding energy $\gapb$ between the fully paired BCS ground 
state and the blocked one of odd--mass nuclei. 
The five--point gap $\gapfive$ provides a reliable quantity to fit the  
effective pairing strength, among all approximations for the pairing gap it  
is closest to $\gapb$. Therefore we use it to calculate the 
experimental pairing gaps and take it as point of reference for all  
other approximations for the pairing gap. Like the other finite--difference 
formulae $\gapfive$ contains a spurious contribution from the mean field
around magic numbers, which is related to the jump of the Fermi energy
at shell closures.

The four--point gap $\gapfour$ has nearly the same overall quality as the 
five--point gap as compared to $\gapb$, but its definition has an ambiguity,  
so that we prefer the five--point gap. The three--point gap $\gapthree$  
has a large contribution from the mean--field, which becomes rather  
obvious looking at proton gaps, consequently this quantity should not be used  
for the fit of the pairing strength since deficiencies of the underlying  
mean--field (especially concerning the symmetry energy) may be visible in the  
$\gapthree$. When comparing calculated and experimental $\gapthree$ 
respectively $\gapfour$ for a well--adjusted interaction, however, both 
gaps show the same quality like the five--point gaps $\gapfive$.

In situations where it is not possible to calculate $\gapfive$, 
e.g.\ looking at potential energy surfaces, a reasonable approximation  
for the pairing gap is provided by \hbox{$\gapuv + \lambda_2$}. By adding  
$\lambda_2$ to the spectral gap the approximate particle--number projection   
gives an improved reproduction of the calculated $\gapfive$ in most cases.  
The spectral gap $\gapuv + \lambda_2$ is in much better agreement  
with the calculated $\gapfive$ than the averaged gap  
$\gapvv + \lambda_2$, which sets too much bias on  
deeply bound states and therefore should not be used in models in which  
the pairing interaction acts mainly at the nuclear surface.  
The single--quasiparticle energies always overestimate the $\gapfive$, 
but this is not too surprising because they are calculated  
from a non--self--consistent wave function of the odd nucleus.

The results show the danger of comparing calculated and experimental gaps  
computed from different definitions. The deviations 
between the various definitions depend on the actual nucleus and become 
generally larger when going towards the drip--lines. 
This is important especially when adjusting the free parameters 
of the pairing interaction. Experimental and calculated gaps should 
be calculated in the same manner, the five--point gap provides a  
useful tool for that.

The choice of the test cases (heavy semi--magic nuclei) and
restriction to spherical nuclei have minimized the impact
of polarization effects on the gaps. Recent explorations of
dynamical polarization effects \cite{Xu99a,uggap} hint that these
may be non--negligible at a quantitative level. Conclusive answers are
yet inhibited due to uncertainties of present mean--field models in the
time--odd channel. This point deserves attention in future work.

Another open question is the functional form of the pairing  
interaction. We find in this paper that the DDDI functional 
gives a slightly better description of pairing gaps compared 
with a delta pairing force, but to give a definitive answer 
one has to look at more nuclei and other data as well. Work 
in this direction is underway.  
% 
%======================================================================= 
% 
\section*{Acknowledgments} 
The authors would like to thank W.~Nazarewicz for  
stimulating discussions and challenging comments. 
This work was supported by Bundesministerium f\"ur Bildung und  
Forschung (BMBF), Project No.\ 06 ER 808, by Gesellschaft f\"ur  
Schwerionenforschung (GSI), by Gra\-du\-ier\-ten\-kolleg Schwerionenphysik 
and by the U.S.\ Department of Energy under Contract   
No.\ DE--FG02--97ER41019 with the University of North Carolina 
and Contract No.\ DE--FG02--96ER40963 with the University of Tennessee
and by the NATO grant SA.5--2--05 (CRG.971541). 
The Joint Institute for Heavy Ion Research has as member institutions the  
University of Tennessee, Vanderbilt University, and the Oak Ridge 
National Laboratory; it is supported by the members and by the Department  
of Energy through Contract No.\ DE--FG05--87ER40361 with the University of  
Tennessee. 
% 
%======================================================================= 
% 
\begin{appendix} 
\section{The calculation of ${\mathbf \lambda_2}$} 
% 
%----------------------------------------------------------------------- 
% 
\subsection{The General Expression} 
\label{Sect:lambda2} 
In case of an underlying many--body Hamiltonian $\hat{H}$ the parameter  
$\lambda_2$ in the variational equation is fixed by the condition 
\begin{equation} 
\label{LN:Nebenbedingungen2} 
\langle \hat{K} \hat{N}_2^2 \rangle  
= 0 
\quad , 
\end{equation} 
where $\hat{K}$ is given by  
\begin{equation} 
\label{eq:Kdef} 
\hat{K} 
=   \hat{H} 
  - \sum_{q \in p,n} \Big(   \lambda_{1,q} \langle \hat{N}_q   \rangle 
                           + \lambda_{2,q} \langle \hat{N}_q^2 \rangle 
                     \Big) 
\end{equation} 
and $\hat{N}_2$ is the two--quasiparticle part of the particle--number 
operator (\ref{eq:N_2}). 
Because we look at like-particle pairing only the equations for 
protons and neutrons separate. Therefore the index $q$ for the isospin  
of the particle--number operator, the single--particle states etc 
can suppressed to get a compact notation. Introducing the shifted  
many--body state \cite{Rei96a} 
\begin{equation} 
| \xi \rangle 
= {\rm e}^{ {\rm i} \xi \hat{N}_2} | 0 \rangle 
\qquad \hbox{with} \qquad  
| 0 \rangle = | \xi \rangle \Big|_{\xi=0} 
\quad , 
\end{equation} 
the condition (\ref{LN:Nebenbedingungen2}) is equivalent to 
\begin{equation} 
\label{LN:Nebenbedingungen3} 
\partial_\xi^2 \langle 0 | \hat{K} | \xi \rangle \Big|_{\xi=0} 
= 0 \quad . 
\end{equation} 
This can be resolved into an expression for $\lambda_2$ 
\begin{equation} 
\label{LN:lambda2} 
\lambda_2 
  =   \frac{\partial_\xi^2 \langle 0 | \hat{H}   | \xi \rangle \Big|_{\xi=0}} 
           {\partial_\xi^2 \langle 0 | \hat{N}^2 | \xi \rangle \Big|_{\xi=0}} 
      \quad . 
\end{equation} 
The denominator of (\ref{LN:lambda2}) can be calculated easily using  
Wick's theorem 
\begin{eqnarray} 
\partial_\xi^2 \langle 0 | \hat{N}^2 | \xi \rangle \Big|_{\xi=0} 
& = &  8 \Big[ \sum_{k \gtrless 0} u_k^2 v_k^2 \Big]^2  
    - 16 \sum_{k \gtrless 0} u_k^4 v_k^4 
\quad . 
\end{eqnarray} 
So far we have formulated the numerator in terms of an underlying 
many--particle Hamiltonian, but the formulation (\ref{LN:Nebenbedingungen3}) 
has the advantage that it can be translated into the formal framework 
of energy functionals \cite{Rei96a} 
\begin{equation} 
\label{LN:L2numerator} 
\langle 0 | \hat{H} | \xi \rangle 
\quad \Rightarrow \quad  
{\cal E}^{(\xi)}  
= {\cal E} [ \hat{\rho}^{(\xi)}, \hat{\chi}^{(\xi)}, \hat{\chi}^{*(\xi)} ] 
\quad . 
\end{equation} 
$\hat{\rho}^{(\xi)}$ and $\hat{\chi}^{(\xi)}$ are the shifted density 
matrix and pair density matrix respectively, which have to be calculated  
now as non-diagonal matrix elements 
\begin{subequations}{} 
\begin{eqnarray} 
\hat{\rho}^{(\xi)} 
& \equiv & \rho^{(\xi)} (\xvec, \xvecp) 
  =        \langle 0 | \hat{\psi}^\dagger (\xvecp) \hat{\psi} (\xvec)  
           | \xi \rangle 
           \\ 
\hat{\chi}^{(\xi)} 
& \equiv & \chi^{(\xi)} (\xvec, \xvecp) 
  =        - \sigma'  
           \langle 0 | \hat{\psi} (\rvecp, -\sigma') \hat{\psi} (\xvec)  
           | \xi \rangle 
           \\ 
\hat{\chi}^{* (\xi)} 
& \equiv & \chi^{* (\xi)} (\xvec, \xvecp) 
  =        - \sigma  
           \langle 0 | \hat{\psi}^\dagger (\xvecp)  
                       \hat{\psi}^\dagger (\rvec, -\sigma)  
           | \xi \rangle 
\end{eqnarray} 
\end{subequations} 
All local densities and currents the Skyrme and pairing energy  
functionals depend on can be derived from these density matrices. 
The second derivative of the energy functional is given by 
\begin{eqnarray} 
\label{eq:d2Edrho} 
\partial_\xi^2 {\cal E}^{(\xi)} 
& = & \trace \bigg\{  
        \frac{\delta \EF}{\delta \hat{\rho}}   \partial^2_\xi \hat{\rho}  
      + \frac{\delta \EF}{\delta \hat{\chi}}   \partial^2_\xi \hat{\chi}  
      + \frac{\delta \EF}{\delta \hat{\chi}^*} \partial^2_\xi \hat{\chi}^*  
      \bigg\} 
      \nonumber \\ 
&   & + \trace \, \trace  
       \bigg\{ \frac{\delta^2 \EF} 
                    {\delta \hat{\rho}_1 \delta \hat{\rho}_2}  
                     \partial_\xi \hat{\rho}_1 \, 
                     \partial_\xi \hat{\rho}_2 
       + 2     \frac{\delta^2 \EF} 
                    {\delta \hat{\chi} \delta \hat{\chi}^*}  
               \partial_\xi \hat{\chi} \, 
               \partial_\xi \hat{\chi}^* 
      \nonumber \\ 
&   & \quad + 2      \frac{\delta^2 \EF} 
                    {\delta \hat{\rho} \, \delta \hat{\chi}}  
               \partial_\xi \hat{\rho} \, 
               \partial_\xi \hat{\chi} 
       + 2     \frac{\delta^2 \EF} 
                    {\delta \hat{\rho} \, \delta \hat{\chi}^*}  
               \partial_\xi \hat{\rho} \, 
               \partial_\xi \hat{\chi}^* 
      \bigg\} 
\; . 
\end{eqnarray} 
$\delta$ stands for a functional derivative. The trace is a shorthand  
notation for the integration and summation over all coordinates 
\begin{equation} 
\trace \left\{ \frac{\delta \EF}{\delta \hat{\rho}} \hat{\rho}  \right\}  
\equiv \iint \! \rmd x \; \rmd x' \;  
       \frac{\delta \EF}{\delta \rho(\xvec,\xvecp)} 
       \rho (\xvec,\xvecp) 
\quad . 
\end{equation} 
The terms with a single trace in (\ref{eq:d2Edrho}) vanish in case of a  
BCS ground state, while the terms with double traces can be  
simplified defining the response density matrices $\tilde{\rho}$,  
$\tilde{\chi}$ and $\tilde{\chi}^*$ 
\begin{equation} 
\tilde{\rho} 
= - \iunit \; \partial_\xi \, \hat{\rho}^{(\xi)} \Big|_{\xi=0} 
\quad , \quad 
\tilde{\chi} 
= - \iunit \; \partial_\xi \, \hat{\chi}^{(\xi)} \Big|_{\xi=0} 
\quad . 
\end{equation} 
leading to the final expression 
\begin{eqnarray} 
\label{LN:L2numerator2} 
\partial^2_\xi {\cal E}^{(\xi)} \Big|_{\xi=0} 
& = & - \trace \, \trace  
      \bigg\{ \frac{\delta^2 \EF} 
                   {\delta \hat{\rho}_1 \delta \hat{\rho}_2}  
              \tilde{\rho}_1 \, 
              \tilde{\rho}_2 
        + 2  
              \frac{\delta^2 \EF} 
                   {\delta \hat{\chi} \delta \hat{\chi}^*}  
              \tilde{\chi} \, 
              \tilde{\chi}^* 
      \nonumber \\ 
&   & \quad + 2 \frac{\delta^2 \EF} 
                   {\delta \hat{\rho} \, \delta \hat{\chi}}  
              \tilde{\rho} \, 
              \tilde{\chi} 
      + 2 \frac{\delta^2 \EF} 
                   {\delta \hat{\rho} \, \delta \hat{\chi}^*}  
              \tilde{\rho} \, 
              \tilde{\chi}^* 
      \bigg\}   
\end{eqnarray} 
which has to be inserted into (\ref{LN:lambda2}).  
Since we are considering pairing between like particles only there are 
no mixed terms with derivatives with respect to proton and neutron densities. 
This expression has to be evaluated now for the pairing and the  
mean--field energy functional. This density--functional approach to the 
calculation of $\lambda_2$ incorporates in a natural way the additional 
contributions to the LN equations arising for density--dependent  
interactions discussed in \cite{Val96a,Val97a}. 
% 
%--------------------------------------------------------------------------- 
% 
\subsection{The Linear Response of a Skyrme Energy Functional} 
\label{Subesct:LRskyrme} 
The Skyrme energy functionals are constructed to be effective interactions  
for nuclear mean--field calculations. For even--even nuclei, the Skyrme 
energy functional used in this paper 
\begin{equation} 
\label{eq:Emf} 
{\cal E}  
  =     {\cal E}_{\rm kin}  
      + {\cal E}_{\rm Sk}   
      + {\cal E}_{\rm C} 
\quad ,  
\end{equation} 
is the sum of the functional of the kinetic energy ${\cal E}_{\rm kin}$, 
the effective functional for the strong interaction ${\cal E}_{\rm Sk}$ 
and the Coulomb interaction ${\cal E}_{\rm C}$ including the exchange term  
in Slater approximation. The actual functionals are given by 
\begin{eqnarray} 
\label{eq:ESkyrme} 
{\cal E}_{\rm kin} 
& = & \frac{\hbar^2}{2m} \int \! {\rm d}^3 r \; \tau 
      \quad , 
      \nonumber \\ 
{\cal E}_{\rm C} 
& = & \frac{e^2}{2} \! \! \iint \! {\rm d}^3 r \, {\rm d}^3 r' 
      \frac{\rho_{\rm p} (\textbf{r}) \rho_{\rm p} (\textbf{r}') } 
           {|\textbf{r} - \textbf{r}'|} 
      - \frac{3 e^2}{4} \! \left( \frac{3}{\pi} \right)^{1/3} \! \! \! \! 
        \int \! {\rm d}^3 r \, \rho_{\rm p}^{4/3} 
      \nonumber \\ 
{\cal E}_{\rm Sk}  
& = & \int \! {\rm d}^3r  
      \bigg[   \frac{b_0}{2} \rho^2 
             + b_1           \rho \tau 
             - \frac{b_2}{2} \rho \Delta \rho  
             + \frac{b_3}{3} \rho^{\alpha +2} 
             - b_4           \rho \nabla \cdot {\bf {J}}  
      \nonumber \\   
&   & \quad - \sum_q \bigg( 
            \frac{b'_0}{2} \rho_q^2 
          + b'_1           \rho_q \tau_q  
          - \frac{b'_2}{2} \rho_q \Delta \rho_q      
      \nonumber \\   
&   & \qquad  \qquad 
          + \frac{b'_3}{3} \rho^\alpha \rho_q^2 
          + b'_4           \rho_q \nabla \cdot {\bf J}_q 
      \bigg) 
      \bigg]  
\quad . 
\end{eqnarray} 
The local density $\rho_q$, kinetic density $\tau_q$ 
and spin--orbit current ${\bf J}_q$ entering the functional are given by 
\begin{eqnarray} 
\rho_q (\vec{r}) 
& = & \sum_{\sigma = \pm}\rho_q (\rvec, \sigma; \rvec, \sigma) 
      \nonumber \\ 
& = & \sum_{k \in \Omega_q} v_k^2 \; | \phi_k (\rvec) |^2 
      \quad , 
      \nonumber \\  
\tau_q (\vec{r}) 
& = & \sum_{\sigma = \pm}  
      \vnabla \cdot \vnabla{}' \rho_q (\rvec, \sigma; \rvecp, \sigma) 
      \Big|_{\rvec = \rvecp} 
      \nonumber \\ 
& = & \sum_{k \in \Omega_q} v_k^2 \; | \nabla \phi_k (\rvec) |^2 
      \quad ,  
      \nonumber \\ 
{\bf J}_q (\vec{r}) 
& = & - {\textstyle \frac{\iunit}{2}} 
      ( \vnabla - \vnabla') \times  
      \sum_{\sigma, \sigma' = \pm} \rho_q (\rvec, \sigma; \rvecp, \sigma') \;  
      \vec{\sigma}_{\sigma' \sigma} 
      \Big|_{\rvec = \rvecp}  
      \nonumber \\ 
& = &  - {\textstyle \frac{\iunit}{2}} \sum_{k \in \Omega_q} v_k^2 \; 
      \Big[ \phi_k^\dagger (\rvec) \, \nabla \times \hat\sigma \,  
            \phi_k (\rvec) 
            - \mbox{ h.c. } 
      \Big] 
\; , 
\end{eqnarray} 
with $q \in \{ {\rm p,n } \}$. $\vec{\sigma}_{\sigma' \sigma}$ is the 
matrix element of the vector of the Pauli spin matrices between the  
unit spinors with spin projection $\sigma'/2$ and $\sigma/2$. 
Densities without index in (\ref{eq:ESkyrme}) denote total densities, 
e.g.\ \mbox{$\rho = \rho_{\rm p} + \rho_{\rm n}$}. The $\phi_k$ are 
the spinors of the single--particle wave functions, the $v^2$ occupation  
probabilities. 
The parameters $b_i$ and $b'_i$ used in the above definition 
are chosen to give a most compact formulation of the energy functional, 
the corresponding mean--field Hamiltonian and residual interaction 
\cite{Rei92}. The Skyrme energy functional contains an extended  
spin--orbit coupling with an explicit isovector degree--of--freedom  
as used in the parameterization SkI4 \cite{Rei95a}.

The response densities needed for the evaluation of  
(\ref{LN:L2numerator2}) are given by 
\begin{eqnarray} 
\tilde\rho_q 
& = & 2 \sum_{k \in \Omega_q } u_k^2 v_k^2 \; 
      | \phi_k |^2 
      \quad ,  
      \nonumber \\ 
\tilde\tau_q  
& = & 2 \sum_{k \in \Omega_q} u_k^2 v_k^2 \; 
      | {\bf \nabla} \phi_k |^2 
      \quad , \nonumber \\ 
\tilde{\bf J}_q 
& = & - \iunit \sum_{k \in \Omega_q} u_k^2 v_k^2 \,  
        \Big[ \phi_k^\dagger \, \nabla \times \hat\sigma \, \phi_k 
            - ( \nabla \times \hat\sigma \, \phi_k )^\dagger \phi_k 
      \Big] \nonumber  
\quad . 
\end{eqnarray} 
Evaluating Eq. (\ref{LN:L2numerator2}) for the Skyrme energy  
functional (\ref{eq:ESkyrme}) leads to 
\begin{eqnarray} 
\label{eq:LN:LRMF1} 
\lefteqn{ 
\partial^2_\xi {\cal E}_{\rm Sk}^{(\xi)} \Big|_{\xi=0} 
} \nonumber \\ 
& = & \trace \; \trace  
      \bigg\{ \tilde{\rho}_{q,1}  
             \frac{\delta^2 {\cal E}_{\rm Sk}} 
                  {\delta \rho_{q,1} \, \delta \rho_{q,2}} \;  
             \tilde{\rho}_{q,2}  
        + 2 \tilde\rho_{q} \;  
       \frac{\delta^2 {\cal E}_{\rm Sk}} 
            {\delta \rho_{q} \, \delta \tau_{q}} \;  
       \tilde \tau_{q} 
       \nonumber \\ 
&   & \qquad + 2 \tilde\rho_{q} \; 
       \frac{\delta^2 {\cal E}_{\rm Sk}} 
            {\delta \rho_{q} \, \delta {\bf J}_{q}} \cdot  
       \tilde{\bf J}_{q}  
       \bigg\} 
      \nonumber \\ 
& = & \int \! \rmd^3r \;  
      \bigg\{ 
          \big( b_0 - b_0^{\prime} \big) \tilde\rho_{q}^2 
      + 2 \big( b_1 - b_1^{\prime} \big) \tilde\rho_{q} \tilde\tau_{q} 
      \\ 
&   & \quad    
      - \big( b_2 - b_2^{\prime} \big) \tilde\rho_{q} \Delta \tilde\rho_{q} 
      - 2 \big( b_4 + b_4^{\prime} \big) \tilde\rho_{q}  
          {\bf \nabla} \cdot \tilde{\bf J}_{q}       
      \phantom{\Big[} 
      \nonumber \\ 
&   &  \quad    
      + \tfrac{1}{3} \big[ (\alpha+2)(\alpha+1) b_3 - 2 b_3' \big] \, 
      \rho^{\alpha} \, \tilde\rho_{q}^2 
      \phantom{\Big[} 
      \nonumber \\ 
&   &  \quad   
      - b_3^{\prime} \tfrac{1}{3}  
        \Big[ 4 \alpha \rho^{\alpha-1} \, \rho_{q} 
              + \alpha (\alpha-1) \rho^{\alpha-2} \sum_{q'} \rho_{q'}^{2} 
        \Big] \, \tilde\rho_{q}^2 
      \bigg\} 
\quad . \nonumber 
\end{eqnarray}       
For the local densities appearing in (\ref{eq:LN:LRMF1}) the trace 
reduces to a spatial integral. 
For the protons one has an additional contribution from the Coulomb 
interaction 
\begin{eqnarray} 
\label{eq:LN:LRMF5}  
\lefteqn{ 
\partial^2_\xi {\cal E}_{\rm C}^{(\xi)} \Big|_{\xi=0} 
} 
      \nonumber \\ 
& = & \trace \, \trace \left\{ 
      \tilde\rho_{\text{p}} \, 
      \frac{\delta^2 {\cal E}_{\text{C}}} 
           {\delta\rho_{\text{p}} \, \delta\rho_{\text{p}}} \;  
      \tilde\rho_{\rm p} \right\} 
      \nonumber \\ 
& = & e^2 \! \! \iint \! \rmd^3r \; \rmd^3r' \,  
      \frac{\tilde\rho_{\text{p}}({\bf r}) \, \tilde\rho_{\text{p}}({\bf r}')} 
           {|{\bf r}-{\bf r}'|}  
      -\frac{e^2}{3} \left( \frac{3}{\pi} \right)^{\! \! 1/3} \! \! 
      \int \! \rmd^3r \, 
      \frac{\tilde\rho_{\text{p}}{}^2}{\rho_{\text{p}}^{2/3}}  
      \nonumber \\ 
& = & \int \! \rmd^3r \; \tilde\rho_{\text{p}} \;  
      \tilde{V}_{\text{coul}} 
      -\frac{e^2}{3} \left( \frac{3}{\pi} \right)^{\! \! 1/3} \! \! 
      \int \!  \rmd^3r \, 
      \frac{\tilde\rho_{\text{p}}{}^2}{\rho_{\text{p}}^{2/3}} 
\quad . 
\end{eqnarray} 
Poisson's equation for the response Coulomb potential  
\begin{equation} 
\Delta \tilde{V}_{\text{coul}} 
= - 4 \pi \; \tilde\rho_{\text{p}}  
\quad . 
\end{equation} 
is solved numerically using the techniques explained in \cite{Coul}. 
The kinetic energy gives no contribution to $\lambda_2$. 
We omit the contribution from the center--of--mass correction  
(and therefore the approximate particle--number correction of  
this term). 
% 
%============================================================================ 
% 
\subsection{The Linear Response of the Pairing Energy Functional} 
For the calculation of the contribution of a pairing energy functional  
of type (\ref{eq:PairFunc}) to Eq.~(\ref{LN:L2numerator2}) the local 
response pair density $\tilde \chi_q$ is needed 
\begin{equation} 
\tilde \chi_q  
  =   - \iunit \; \partial_\xi \chi_q^{(\xi)} \Big|_{\xi = 0} 
  =   - 4 \sum_{k \in \Omega_q \atop k > 0}  
      f_k^2 \; u_k^3 v_k \; | \phi_k |^2 
\quad . 
\end{equation} 
The response pair density is not hermitian, the adjoint response 
pair density $\tilde\chi^{\ast}_q$ reads 
\begin{equation} 
\tilde\chi^{\ast}_q 
= - \iunit \; \partial_\xi \chi_q^{\ast (\xi)} \Big|_{\xi = 0} 
= 4 \sum_{k \in \Omega_q \atop k > 0} 
  f_k^2 \; u_k v_k^3 \; | \phi_k |^2 
\quad . 
\end{equation} 
In case of the DF pairing energy functional there is only one contribution 
from the derivatives with respect to the pair density (\ref{LN:L2numerator2}) 
\begin{eqnarray} 
\label{eq:LN:LRPair1} 
\partial^2_\xi {\cal E}_{\rm DF}^{(\xi)} \Big|_{\xi=0} 
& = & 2 \, \trace \, \trace \left\{ 
      \tilde{\chi}_{q} \;  
      \frac{\delta^2 {\cal E}_{\rm DF}} 
           {\delta \chi_{q} \, \delta \chi^\ast_{q}} \;  
      \tilde{\chi}^\ast_{q} \right\} 
      \nonumber \\  
& = &  
      \frac{V_q}{2} \int \! \rmd^3r \;  
      \tilde\chi_q^\ast \; \tilde\chi_q 
      \quad . 
\end{eqnarray} 
In case of the DDDI pairing energy functional there are additional  
contributions from derivatives with respect to the local density 
\begin{eqnarray} 
\label{eq:LN:LRPair2} 
\lefteqn{ 
\partial^2_\xi {\cal E}_{\rm DDDI}^{(\xi)} \Big|_{\xi=0} 
} \nonumber \\ 
& = & \trace \, \trace  
      \bigg\{ 2 \tilde{\chi}_{q} \, 
              \frac{\delta^2 {\cal E}_{\text{\rm DDDI}}} 
                   {\delta \chi_{q} \, \delta \chi^\ast_{q}} \,  
              \tilde{\chi}^\ast_{q} 
         + 2  \tilde{\rho}_{q} \, 
              \frac{\delta^2 {\cal E}_{\text{\rm DDDI}}} 
                   {\delta \rho_{q} \, \delta \chi_{q}} 
              \, \tilde{\chi}_{q} 
      \nonumber \\  
&   & \quad  
      + 2 \tilde{\rho}_q \, 
              \frac{\delta^2 {\cal E}_{\text{\rm DDDI}}} 
                   {\delta \rho_{q} \, \delta \chi^\ast_{q}} 
              \, \tilde{\chi}^\ast_{q} 
      +  \tilde{\rho}_{q,1} \, 
              \frac{\delta^2 {\cal E}_{\text{\rm DDDI}}} 
                   {\delta \rho_{q,1} \, \delta \rho_{q,2}} \, 
              \tilde{\rho}_{q,2} 
      \bigg\} 
      \nonumber \\ 
& = & \frac{V_q}{2} \int \! \rmd^3r \;  
      \bigg\{   \tilde\chi_q^\ast  \; \tilde\chi_q  
                \Big[ 1 - \left( {\textstyle \frac{\rho}{\rho_0} }  
                          \right)^\gamma  
                \Big] 
      \nonumber \\  
&   & \quad                      
             - \frac{\gamma}{\rho_0^\gamma} \;  
               \tilde\rho_q \; \rho^{\gamma-1} 
                \Big(  \tilde\chi^\ast_q \; \chi_q 
                      + \chi^\ast_q \; \tilde\chi_q 
                \Big) 
      \nonumber \\  
&   & \quad                      
             - \frac{\gamma (\gamma-1) }{2 \rho_0^\gamma} 
               \tilde\rho_q^2 \; \rho^{\gamma-2} 
               \chi^\ast_q \; \chi_q 
      \bigg\}  
\quad . 
\end{eqnarray} 
% 
%======================================================================= 
% 
\section{Single--Quasiparticle Energies} 
\label{Sect:OQPE} 
The single--quasiparticle energies are deduced from the 
non--self--consistent \emph{ansatz}  
\begin{equation} 
\label{eq:1qps} 
\ket{k} 
= \alopk_k \, \kBCS 
= v_k \, \aopk_k 
  \prod_{m > 0 \atop m \neq k} ( u_m + v_m \aopk_m \aopk_{\bar{m}} )  
  \; \kvac 
\end{equation} 
for the blocked many--body wave function of the odd mass number  
nucleus \cite{EGIII,Ringbook,Nilssonbook}.  
The single--particle wave functions of all states and the occupation 
probabilities of all unblocked states \mbox{$m \neq k$} are taken from 
the self--consistent calculation of the BCS ground state. 
The normalization of $\ket{k}$ requires \mbox{$v_k = 1$}, 
\mbox{$u_k = v_k = v_{\bar{k}} = 0$}. Note that $\ket{k}$ does 
not yield the proper particle number of the excited state because the 
occupation of the blocked pair of states $(k,\bar{k})$ in the  
BCS ground state in general will differ from one. 
 
The excitation energy $E_k$ of the lowest one--quasi\-par\-ticle state is an  
approximation for the odd--even mass difference. To take the difference 
in particle number between the fully paired BCS state and $\ket{k}$  
into account, $E_k$ has to be calculated from the difference of  
${\cal K}$ given by (\ref{eq:calK}) calculated for $\ket{k}$ and the  
BCS ground state. This leads to 
\begin{equation} 
E_k 
=   {\cal E}^{(k)}  
  + \bra{k} \hat{N}' \ket{k} 
  - {\cal E}^{(0)} - \langle \hat{N}' \rangle 
\end{equation} 
where we have introduced the abbreviation 
\begin{equation} 
\hat{N}'  
= - \sum_{q \in \{ {\rm p,n} \}} \Big(   \lambda_{1,q} \Nop_q 
                                       + \lambda_{2,q} \Nop^2_q \Big) 
\quad . 
\end{equation} 
${\cal E}^{(0)}$ is the energy functional of the BCS ground state, 
while ${\cal E}^{(k)}$ is the energy functional evaluated for the  
one--quasiparticle state $\ket{k}$.  
 
The expectation values of $\hat{N}'$ are conveniently calculated  
with standard operator techniques \cite{EGIII,Ringbook,Nilssonbook} 
\begin{equation} 
\bra{k} \Nop' \ket{k} 
= \langle \alop_k \Nop' \; \alopk_k \rangle 
= \Nop_{00}' + \langle \alop_k \Nop_{11}' \alopk_k \rangle 
\quad , 
\end{equation} 
where $\Nop_{00}'$ is the BCS expectation value of $\hat{N}'$ while  
$\Nop_{11}'$ is its (11) component. 
The Bogolyubov transformation of the particle--number operator 
reads 
\begin{eqnarray} 
\label{eq:Nop} 
\Nop  
& = & \sum_{k \gtrless 0} \aopk_k \aop_k 
      \nonumber \\ 
& = & \sum_{k \gtrless 0} 
      \Big[ 
        v_k^2  
      + ( u_k^2 - v_k^2 ) \alopk_k \alop_k 
      + u_k v_k ( \alopk_k \alopk_{\bar k}  
      +           \alop_{\bar k} \alop_k ) 
      \Big] 
      \nonumber \\ 
& = & \Nop_{00} + \Nop_{11} + \Nop_{20} + \Nop_{02} 
\quad . 
\end{eqnarray} 
$(\Nop^2)_{11}$ can be calculated from (\ref{eq:Nop}) without 
repetition of the quasiparticle transformation. One obtains 
\begin{equation} 
\label{eq:Nopsqr11} 
(\Nop^2)_{11} 
= \sum_{k \gtrless 0} 
  \Big\{ ( u_k^2 - v_k^2 ) [ 2 (N + 1) - 4 v_k^2 ] - 1  
  \Big\} \alopk_k \alop_k 
\quad . 
\end{equation} 
The calculation of the contribution of a many--body Hamiltonian operator  
to $E_k$ can be found in many textbooks, see e.g.\  
\cite{EGIII,Ringbook,Nilssonbook}. Here, however, we want to calculate it  
in the framework of effective energy functionals. The energy functional  
${\cal E}^{(k)} = {\cal E}[\hat\rho^{(k)}, \chi^{(k)}, \chi^{* (k)}]$ 
of the one--quasiparticle state is to be calculated from the density  
matrix and pair density matrix evaluated for the one--quasiparticle  
state $\ket{k}$ 
\begin{subequations}{} 
\begin{eqnarray} 
\lefteqn{ 
\rho^{(k)}(\xvec; \xvecp) 
  =   \bra{k} \hat{\psi}^\dagger(\xvecp) \hat{\psi}(\xvec) 
      \ket{k}  
      \phantom{ \sum_{m > 0}} 
}     \nonumber \\ 
& = &  \phi^{*}_k (\xvecp) \phi_k (\xvec)  
       + \sum_{m > 0\atop m \neq k} v_m^2 \;  
         \phi^{*}_m (\xvecp) \, \phi_m (\xvec)    
      \quad , \\ 
\lefteqn{ 
\chi^{(k)} (\vec{r}) 
  =   - \sum_{\sigma = \pm} \sigma  
      \bra{k} \hat{\psi}(\vec{r},-\sigma) 
              \hat{\psi}(\xvec) 
      \ket{k} 
}     \nonumber \\ 
& = & -2 \sum_{m > 0 \atop m \neq k} f_m u_m v_m  
      | \phi(\vec{r})|^2 
      \quad . 
\end{eqnarray} 
\end{subequations} 
$\rho^{(k)}(\vec{x}; \vec{x}')$ and $\chi^{(k)} (\vec{r})$ 
are inserted into the definition of the energy functionals 
(\ref{eq:Emf}) and (\ref{eq:PairFunc}). This leads to  
\begin{equation} 
E_k 
= {\cal E}^{(k)} - {\cal E}^{(0)} 
  - (\lambda_{q} - 4 \lambda_{2,q} v_k^2) (u_k^2 - v_k^2) + \lambda_{2,q} 
\end{equation} 
where $\lambda_q = \lambda_{1,q} + 4 \lambda_{2,q} (N_q + 1)$ was used. 
 
The usual approximation for the one--quasiparticle energy is obtained 
taking only the linear change in the densities into account 
\begin{equation} 
{\cal E}^{(k)} 
\approx {\cal E}^{(0)} 
        + {\rm tr} \left\{  
          \frac{\delta {\cal E}}{\delta \hat\rho} \delta \hat\rho 
          + \frac{1}{2} \frac{\delta {\cal E}}{\delta \chi} \delta \chi 
          + \frac{1}{2} \frac{\delta {\cal E}}{\delta \chi^*} \delta \chi^* 
        \right\} 
\end{equation} 
with 
\begin{subequations}{} 
\begin{eqnarray} 
\delta \hat\rho 
& = & \hat\rho^{(k)} - \hat\rho 
  =   (1 - 2 v_k^2) \; \phi^*_k (\xvecp) \; \phi_k (\xvec) 
      \quad , \\ 
\delta \chi 
& = & \delta \chi^* 
  =   \chi^{(k)} - \chi 
  =   2 \, u_k v_k \; | \phi_k |^2 
      \quad . 
\end{eqnarray} 
\end{subequations} 
This yields 
\begin{equation} 
{\cal E}^{(k)} - {\cal E}^{(0)} 
\approx \epsilon_k (u_k^2 - v_k^2) + 2 f_k \Delta_k u_k v_k 
\quad . 
\end{equation} 
Together with the contributions from the particle--number operators, 
Eqns.~(\ref{eq:Nop}) and (\ref{eq:Nopsqr11}), one obtains 
\begin{eqnarray} 
\label{Gap:H11EWa} 
E_k 
& = & ( \epsilon_k' - \lambda_q ) \; ( u_k^2 - v_k^2 ) 
      + 2 f_k \Delta_{k} \; u_k v_k 
      + \lambda_2 
      \nonumber \\ 
& = & \sqrt{ ( \epsilon'_k - \lambda_q )^2 + f_k^2 \Delta_{k}^2 } 
      + \lambda_{2,q} 
\quad , 
\end{eqnarray} 
where the definitions of the Fermi energy (\ref{eq:Efermi}), the  
renormalized single--particle energy (\ref{eq:Erenorm}) and  
the expressions for $v_k^2$ and $u_k^2 = 1 - v_k^2$ have been inserted.  
 
\end{appendix} 
% 
%======================================================================= 
% 

% 
%======================================================================= 
%
\end{document}